\documentclass{jfm}

\usepackage{amsmath}
\usepackage{amssymb}
\usepackage{graphicx}
\usepackage{subcaption}
\usepackage{tikz}
\usepackage{epstopdf, epsfig}
\usepackage{booktabs}
\usepackage{bm}
\usepackage{xr}
\usepackage{pgfplots}
\usepackage{pdfpages}
\allowdisplaybreaks

\newcommand{\pardiff}[2]{\frac{\partial #1}{\partial #2}}

\begin{document}

%\twocolumn[{
%\begin{frontmatter}

\title[Free-surface flow over bottom topography]{Three dimensional free-surface flow over arbitrary bottom topography}

%\author[Scott]{Nicholas Buttle}
%
%\author[Scott]{Ravindra Pethiyagoda}
%
%\author[Scott]{Scott W. McCue\corref{cor1}}
%\ead{scott.mccue@qut.edu.au}
%
%\author[Scott]{Timothy J. Moroney}
%
%\cortext[cor1]{Corresponding author: Tel: +61 (0)7 31384295}
%
%\address[Scott]{School of Mathematical Sciences, Queensland University of %Technology, Brisbane QLD 4001, Australia}

\author[N. R. Buttle, R. Pethiyagoda, T. J. Moroney and S. W. McCue]%
{\small Nicholas R. Buttle, Ravindra Pethiyagoda, Timothy J. Moroney \\ and Scott W. McCue\thanks{Email address for correspondence: scott.mccue@qut.edu.au}}

% NOTE: A full address must be provided: department, university/institution, town/city, zipcode/postcode, country.
\affiliation{\small School of Mathematical Sciences, Queensland University of Technology, Brisbane QLD 4001, Australia}

\maketitle

\begin{abstract}
We consider steady nonlinear free surface flow past an arbitrary bottom topography in three dimensions, concentrating on the shape of the wave pattern that forms on the surface of the fluid.  Assuming ideal fluid flow, the problem is formulated using a boundary integral method and discretised to produce a nonlinear system of algebraic equations.  The Jacobian of this system is dense due to integrals being evaluated over the entire free surface. To overcome the computational difficulty and large memory requirements, a Jacobian-free Newton Krylov (JFNK) method is utilised. Using a block-banded approximation of the Jacobian from the linearised system as a preconditioner for the JFNK scheme, we find significant reductions in computational time and memory required for generating numerical solutions. These improvements also allow for a larger number of mesh points over the free surface and the bottom topography. We present a range of numerical solutions for both subcritical and supercritical regimes, and for a variety of bottom configurations.  We discuss nonlinear features of the wave patterns as well as their relationship to ship wakes.
\end{abstract}

%\begin{keywords}
%surface gravity waves; boundary integral method; ship wakes; Jacobian-free Newton-Krylov method
%\end{keywords}

%\end{frontmatter}
%}]

\section{Introduction}

This paper is concerned with computing three-dimensional steady free-surface flow of an inviscid fluid over a bottom topography.  Upstream from the bottom disturbance, the flow approaches an undisturbed free stream, while downstream from the disturbance there is a wave pattern on the surface which depends on the features of the bottom topography and the speed of the oncoming flow.  The surface wave pattern is closely related to a ship wake generated at the stern of a steadily moving vessel.  Other realisations of these flows include large scale oceanic flows over seamounts or rivers flowing steadily over underwater ridges and isolated obstructions.

The two-dimensional version of this problem has an extensive history.  Typically, complex variable methods are used to reformulate the problem in terms of a integral equation, which can be solved numerically using collocation \citep{Binder2013,Binder2006,Chuang2000,Dias2002,Forbes1982,Hocking2013,King1990,pethiyagoda17c,Zhang1996}.  With these schemes, the effects of bottom topography and nonlinearity on the downstream waves can be explored in some detail.  Alternatively, other approaches include solving for the coefficients in a series solution \citep{Dias1989,Higgins2006}. In contrast to fully numerical studies, analytical progress has been made in two dimensions by linearising the problem (limit of vanishingly small bottom obstruction) \citep{Forbes1982,Gazdar1973,King1990,Lamb1932}, applying a weakly nonlinear expansion which leads to a steady forced Korteweg de Vries equation \citep{Binder2014,Binder2006,Chardard2011,Dias2002}, or using asymptotic analysis (limit of vanishingly small fluid speed) \citep{Chapman2006,Lustri2012}.

Steady flow in three dimensions presents a number of additional challenges.  First, complex variable methods no longer apply in three dimensions, thus many elegant mathematical tricks are no longer appropriate.  Second, even if a boundary integral approach is used, the numerical task is much more demanding due to the size of the problem.  Third, the actual wave patterns that develop behind a bottom obstruction are much more complicated in three dimensions.  For these reasons, despite the extensive history of studying the two-dimensional version (see above), there are surprisingly few attempts to analyse the three-dimensional problem.  Certainly, the linear problem is reasonably straightforward to handle with Fourier transforms (many such problems are treated in \citet{wehausen60}, for example) and there is a history of attention given to various three-dimensional problems involving internal gravity waves (mountain waves) in compressible fluids  \citep{Broutman2003,Eckermann2010,Teixeira2014}.   However, as far as we are aware, there has been no thorough numerical study of fully nonlinear steady free-surface flows of an ideal fluid over a bottom topography that give rise to surface gravity wave patterns.

In this paper, we start by following the approach of \citet{Forbes1} and others \citep{Parau1,Parau2011,Parau2005,Parau2,Parau2007a,Parau2007b} who apply a boundary integral formulation based on Green's functions for three-dimensional steady flows past submerged singularities and surface pressure distributions.  The effect of this framework is to reduce a three-dimensional problem into a two-dimensional one.  We adapt their formulation to apply for flow past a bottom obstruction and then build on the work of \citet{Pethiyagoda14,pethiyagoda2014} to develop a Jacobian Free Newton Krylov method to solve the nonlinear system arising from collocation.  The problems considered in \citet{Pethiyagoda14,pethiyagoda2014} are for an infinitely deep fluid and so do not take into account a channel bottom which we assume is of arbitrary shape.  As such, the preconditioner required in~\citet{Pethiyagoda14,pethiyagoda2014} has a more simple structure with only one fully dense block.  On the other hand, for our finite-depth flows, we have had to take a more considered approach in terms of blocking and banding the preconditioner.  We find that our numerical scheme is highly effective and allows the much larger number of grid points when compared to the standard approach without preconditioning.

The outline of our paper is as follows.  In \S\ref{sec:formulation} we formulate the problem in terms of a velocity potential $\Phi(x,y,z)$, which satisfied Laplace's equation throughout the fluid domain, and the unknown free surface $z=\zeta(x,y)$.  There are two inputs to the dimensionless problem, the Froude number $F$ and the dimensionless description of the bottom topography, $z=\beta(x,y)$.  We recast the problem by applying a boundary integral method so that ultimately the governing equations are a system of two integrodifferential equations.  In \S\ref{sec:numerical} we discretise our governing equations to produce a nonlinear system of algebraic equations, which we solve using a Jacobian-free Newton-Krylov method.  The preconditioner we use is related to the linear problem, as we explain in \S\ref{sec:linear}.  In \S\ref{sec:results} we present numerical results, including free-surface profiles for subcritical and supercritical flows and a variety of bottom configurations.  Our results highlight the role of nonlinearity as we illustrate features of surface wave patterns that are observed only for highly nonlinear solutions.  Finally, we close our paper in \S\ref{sec:discussion} with a summary and discussion of the relationship between our solutions and those due to steadily moving ships.

\section{Formulating the problem}\label{sec:formulation}

\subsection{Problem definition}

\begin{figure}
\centering
\includegraphics[width=0.5\textwidth]{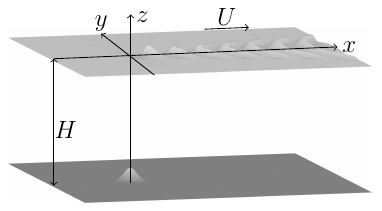}
\caption{Diagram of the flow problem. The $x$-axis is aligned with the direction of the flow far upstream. The fluid is bounded above by the free surface $z=\zeta(x,y)$ and bounded below by the topography $z=\beta(x,y)$ which, in this schematic, is an isolated bump.}
\label{fig:topography_schematic}
\end{figure}

We consider the irrotational flow of an inviscid, incompressible fluid bounded above by a free surface and below by a channel bottom with arbitrary topography.  Far upstream, there is a uniform stream flowing with constant velocity $U$ and depth $H$.  Cartesian coordinates are set up such that the flow is predominantly in the positive $x$-direction and gravity acts the negative $z$-direction.  Due to the bottom topography, a wave pattern resembling a ship's wake appears on the free surface. A schematic of the problem set-up is shown in Figure~\ref{fig:topography_schematic}.  We are interested in the steady-state problem that arises from the long-time limit of this set-up.

The problem is non-dimensionalised by scaling all lengths by the upstream depth, $H$, and all velocities by $U$. By denoting the free surface by $z=\zeta(x,y)$ and the bottom boundary by $z=\beta(x,y)$, the dimensionless problem is to solve Laplace's equation for the velocity potential $\Phi(x,y,z)$,
\begin{equation}
\nabla^{2}\Phi = \pardiff{\Phi^2}{x^2} + \pardiff{\Phi^2}{y^2} + \pardiff{\Phi^2}{z^2} = 0, \quad \beta(x,y) < z < \zeta(x,y), \label{eq:laplace}
\end{equation}
subject to the kinematic, dynamic and radiation conditions:
\begin{align}
\Phi_{x}\zeta_{x} + \Phi_{y}\zeta_{y} = \Phi_{z}, \quad &z = \zeta(x,y), \label{eq:new_finite_kinematic1} \\
\Phi_{x}\beta_{x} + \Phi_{y}\beta_{y} = \Phi_{z}, \quad &z = \beta(x,y), \label{eq:new_finite_kinematic2} \\
\frac{1}{2}\left(\Phi_{x}^{2} + \Phi_{y}^{2} + \Phi_{z}^{2}\right) + \frac{\zeta}{F^{2}} = \frac{1}{2}, \quad &z = \zeta(x,y), \label{eq:finite_dynamic} \\
(\Phi_{x},\Phi_{y},\Phi_{z}) \rightarrow (1,0,0), \quad & x \rightarrow -\infty, \label{eq:new_finite_farfield1} \\
\zeta \rightarrow 0, \quad &x \rightarrow -\infty, \label{eq:new_finite_farfield3}
\end{align}
where $F$ is the depth-based Froude number, $F = U / \sqrt{gH}$. We assume the dimensionless bottom topography $z=\beta(x,y)$ has the property $\beta \rightarrow -1$ as $x^{2}+y^{2} \rightarrow \infty$; that is, the bottom surface becomes flat away from the origin.

The velocity potential on the free surface $z=\zeta(x,y)$ is denoted by $\phi(x,y)=\Phi(x,y,\zeta(x,y))$; on the bottom boundary $z=\beta(x,y)$, it is denoted by $\psi(x,y)=\Phi(x,y,\beta(x,y))$.  With this notation, the dynamic condition \eqref{eq:finite_dynamic} can be rewritten to incorporate the kinematic condition \eqref{eq:new_finite_kinematic1} as:
\begin{align}
\frac{1}{2}\frac{(1+\zeta_{x}^{2})\phi_{y}^{2} + (1+\zeta_{y}^{2})\phi_{x}^{2} - 2\zeta_{x}\zeta_{y}\phi_{x}\phi_{y}}{1 + \zeta_{x}^{2} + \zeta_{y}^{2}} + \frac{\zeta}{F^{2}} = \frac{1}{2}, \quad &z = \zeta(x,y). \label{eq:new_finite_dynamic}
\end{align}
We use this form of the boundary condition below.

{ At this stage it is worth noting that once a solution is computed, the pressure acting on the channel bottom can be calculated by rearranging and evaluating Bernoulli's equation along the bottom boundary to give
\begin{equation*}
p(x,y,\beta) =\frac{1}{2} -\frac{1}{2}\frac{(1+\beta_{x}^{2})\psi_{y}^{2} + (1+\beta_{y}^{2})\psi_{x}^{2} - 2\beta_{x}\beta_{y}\psi_{x}\psi_{y}}{1 + \beta_{x}^{2} + \beta_{y}^{2}} - \frac{\beta}{F^{2}}.
\label{eq:bottompressure}
\end{equation*}
Our numerical scheme presented in Section~\ref{sec:numerical} solves for the velocity potential $\psi$.  As such, the pressure field $p(x,y,\beta)$ can be computed at very little additional cost and thus, in principle, the total force acting on the bottom topography can be calculated numerically by integrating the pressure over the bottom boundary.}

To keep the scheme general, the bottom topography is left arbitrary in our formulation.  However, for the purposes of presenting results, we make use of the bottom shape
\begin{align}
\beta(x,y) = -1+\epsilon\, \mathrm{e}^{-\left({\left(x-b_{1}\right)^{2} + \left(y-b_{2}\right)^{2}}\right)/{2\delta^2}},
\label{eq:bottom}
\end{align}
which represents a Gaussian bump centred at $(b_1,b_2)$ with height $\epsilon$ and ``width'' $\delta$. We shall also use combinations of Gaussian bumps with different centres and widths.

\subsection{Boundary integral formulation}

In order to compute numerical solutions to \eqref{eq:laplace} subject to \eqref{eq:new_finite_kinematic1}-\eqref{eq:new_finite_farfield3}, the problem is now reformulated into an integral equation using Green's second theorem on the functions $\Phi - x$ in the flow domain, with the Green's function
\begin{equation*}
\frac{1}{R_{PQ}} = \frac{1}{4\pi}\frac{1}{((x-x^*)^2 + (y-y^*)^2 + (z-z^*)^2)^{1/2}}.
\end{equation*}
Omitting the details, this reformulation results in the integral equation
\begin{align}
2\pi\left(\phi(x^{*},y^{*}) - x^{*}\right) = I_{1} + I_{2} + I_{3} + I_{4}, \label{eq:new_finite_boundary_integral}
\end{align}
where
\begin{align*}
I_{1} &= \iint_{\mathbb{R}^{2}} \left(\phi(x,y)-\phi\left(x^{*},y^{*}\right)-x+x^{*}\right) \nonumber \\
			& \qquad \times K_{1}(x,y,x^{*},y^{*},\zeta,\zeta_{x},\zeta_{y},\zeta(x^*,y^*)) \text{ d}x\text{ d}y, \\
I_{2} &= \iint_{\mathbb{R}^{2}} \zeta_{x}(x,y)K_{2}(x,y,x^{*},y^{*},\zeta,\zeta_{x},\zeta_{y},\zeta(x^*,y^*)) \text{ d}x\text{ d}y, \\
I_{3} &= \iint_{\mathbb{R}^{2}} \left(\psi(x,y)-x\right) \nonumber \\
			& \qquad \times -K_{1}(x,y,x^*,y^*,\beta,\beta_{x},\beta_{y},\zeta(x^*,y^*)) \text{ d}x\text{ d}y, \\
I_{4} &= \iint_{\mathbb{R}^{2}} \beta_{x}(x,y)K_{2}(x,y,x^{*},y^{*},\beta,\beta_{x},\beta_{y},\zeta(x^*,y^*)) \text{ d}x\text{ d}y,
\end{align*}

\begin{align*}
K_{1}(x,y,x^{*},y^{*},a,b,c,d) &= \frac{a - d - (x-x^{*})b - (y-y^{*})c}{\left((x-x^*)^2 + (y-y^*)^2 + (a-d)^2\right)^{\frac{3}{2}}}, \\
K_{2}(x,y,x^{*},y^{*},a,b,c,d) &= \frac{1}{\sqrt{(x-x^{*})^{2} + (y-y^{*})^{2} + (a-d)^{2}}}.
\end{align*}

Very similar boundary integral formulations have been applied by other authors for steady three-dimensional flows, for example for flows involving submerged singularities~\citep{Forbes1,Forbes2005,Pethiyagoda14,pethiyagoda2014} and flows past pressure distributions \citep{Parau1,Parau2011,Parau2005,Parau2,Parau2007a,Parau2007b}.  The key difference here is that we have allowed for an arbitrary bottom topography, which makes (\ref{eq:new_finite_boundary_integral}) more complicated.

In summary, we have reformulated our problem with the use of a boundary integral method so that, for a given bottom topography $\beta(x,y)$ and Froude number $F$, we are required to solve the integral equation (\ref{eq:new_finite_boundary_integral}) and the dynamic condition (\ref{eq:new_finite_dynamic}) for the two velocity potentials $\phi(x,y)$, $\psi(x,y)$ and the surface elevation $\zeta(x,y)$.  For boundary conditions we impose the radiation condition (\ref{eq:new_finite_farfield1})-(\ref{eq:new_finite_farfield3}).

\section{Numerical method}\label{sec:numerical}
\subsection{Collocation scheme}\label{sec:numerical_collocation}

The governing equations (\ref{eq:new_finite_dynamic}) and (\ref{eq:new_finite_boundary_integral}) are highly nonlinear.  For the most gentle case in which the bottom topography is almost flat, we can simplify these equations to produce a linear problem, as detailed in \S\ref{sec:linear}.  However, to capture the full nonlinear features that arise for distinctively non-flat bottom topographies, we solve the equations numerically.  With that in mind, the infinite plane is truncated such that $-\infty<x<\infty$ becomes $x_{1}<x<x_{N}$ and $-\infty<y<\infty$ becomes $y_{1}<y<y_{M}$. Mesh points are then introduced as $x_{i} = x_{1}+(i-1)\Delta x$, $i=1,2,...,N$ and $y_{j} = y_{1}+(j-1)\Delta y$, $j=1,2,...,M$. The point $x_{1}$ is chosen as a suitably far-upstream value such that the surface is considered flat there. The point $y_{1}$ is chosen such that $y_{1} = -y_{M}$, although the scheme allows for asymmetrical meshes. The vector of $3(N+1)M$ unknowns $\mathbf{u}$ is:
\begin{multline}
\mathbf{u} = \left[ \zeta_{1,1}, (\zeta_{x})_{1,1}, (\zeta_{x})_{2,1}, \ldots, (\zeta_{x})_{N,1}, \zeta_{1,2}, (\zeta_{x})_{1,2}, \ldots, (\zeta_{x})_{N,M}, \right. \\
										\left. \phi_{1,1}, (\phi_{x})_{1,1}, \ldots, (\phi_{x})_{N,M}, \psi_{1,1}, (\psi_{x})_{1,1}, \ldots, (\psi_{x})_{N,M} \right].
\label{eq:unknowns}
\end{multline}
The components of $\mathbf{u}$ are ordered by slicing the domain in the $x$-direction. The values of the functions at the upstream truncation are placed before the derivatives of the corresponding slice.

The problem is solved using an inexact Newton method with an equation of the form $\mathbf{F}(\mathbf{u})=\mathbf{0}$. First, the values for the equation for the free surface and the velocity potential on the free surface and the bottom surface are calculated using the trapezoidal rule. The singular integral $I_{2}$ can be written as $I_{2} = I_{2}' + I_{2}''$, where
\begin{align*}
I_{2}' &= \int_{y_{1}}^{y_{M}} \int_{x_{1}}^{x_{N}} \left(\zeta_{x}(x,y)K_{2} - \zeta_{x}\left(x^{*},y^{*}\right)S_{2}\right) \; \text{d}x\text{d}y, \\
I_{2}'' &= \zeta_{x}\left(x^{*},y^{*}\right) \int_{y_{1}}^{y_{M}} \int_{x_{1}}^{x_{N}} S_{2} \; \text{d}x\text{ d}y,
\end{align*}
with
\begin{align*}
S_{2} &= \frac{1}{(A(x - x^{*})^{2} + B(x - x^{*})(y - y^{*}) + C(y - y^{*})^{2})^{1/2}},
\end{align*}
and
\begin{align*}
A &= 1 + \zeta_{x}^{2}(x^{*},y^{*}), & B &= 2\zeta_{x}(x^{*},y^{*})\zeta_{y}(x^{*},y^{*}), & C &= 1 + \zeta_{y}^{2}(x^{*},y^{*}).
\end{align*}
The integral $I_{2}''$ can be evaluated analytically to give
\begin{align*}
 \int \int \frac{\text{d}s\text{ d}t}{\sqrt{As^{2}+Bst+Ct^{2}}} &= \frac{t}{\sqrt{A}}\ln\left(2As + Bt + 2\sqrt{A(As^{2}+Bst+Ct^{2})}\right) \\
& \quad + \frac{s}{\sqrt{C}}\ln\left(2Ct + Bs + 2\sqrt{C(As^{2} + Bst + Ct^{2})}\right)
\end{align*}

Evaluating the boundary integral equation \eqref{eq:new_finite_boundary_integral} at the half mesh points $(x_{i+\frac{1}{2}},y_{j}$, $i=1,2,...,N-1$, $j=1,2,...,M$ produces $(N-1)M$ equations. Evaluating the boundary integral equation \eqref{eq:new_finite_boundary_integral} at the half mesh points on the bottom surface, that is replacing $\zeta(x^*,y^*)$ with $\beta(x^*,y^*)$ and $\phi(x^*,y^*)$ with $\psi(x^*,y^*)$, produces another $(N-1)M$ equations. A further $(N-1)M$ equations come from evaluating the dynamic condition \eqref{eq:new_finite_dynamic} at the half mesh points. The last $6M$ equations come from the radiation condition outlined in \citet{Scullen98},
\begin{align}
x_{1}(\zeta_{x})_{1,j} + n\zeta_{1,j} &= 0, \label{eq:scullen1}\\
x_{1}(\zeta_{xx})_{1,j} + n(\zeta_{x})_{1,j} &= 0, \\
x_{1}((\phi_{x})_{1,j} - 1) + n(\phi_{1,j} - x) &= 0, \\
x_{1}(\phi_{xx})_{1,j} + n((\phi_{x})_{1,j} - 1) &= 0, \\
x_{1}(\psi_{x})_{1,j} - 1) + n(\psi_{1,j} - x) &= 0, \\
x_{1}(\psi_{xx})_{1,j} + n((\psi_{x})_{1,j} - 1) &= 0, \label{eq:scullen6}
\end{align}
for $j=1,2,...,M$. The value $n=0.05$ represents the upstream decay rate; this particular value was also used by \citet{Pethiyagoda14}. We now have $3(N+1)M$ nonlinear equations (that make up the vector $\mathbf{F}$) for the $3(N+1)M$ unknowns in (\ref{eq:unknowns}).

\subsection{Jacobian-free Newton-Krylov method}
A Jacobian-free Newton-Krylov method is used to solve the system
\begin{align*}
\mathbf{F}(\mathbf{u}) &= \mathbf{0}.% \label{eq:numerical_eq}.
\end{align*}
This method uses a damped Newton iteration
\begin{align*}
\mathbf{u}_{k+1} &= \mathbf{u}_{k} + \mu_{k}\delta\mathbf{u}_{k}, %\label{eq:newton_step}
\end{align*}
with a simple line search to find $\mu_{k}$ such that there is a sufficient decrease in the nonlinear residual. The Newton step is found by satisfying
\begin{align*}
\mathbf{J}(\mathbf{u}_{k})\delta\mathbf{u}_{k} &= -\mathbf{F}(\mathbf{u}_{k}), %\label{eq:newton_linear}
\end{align*}
where $\mathbf{J}$ is the Jacobian matrix. A Generalised Minimum Residual (GMRES) method \citep{Saad1986} with right preconditioning is used to solve for $\delta\mathbf{u}_{k}$. The approximate solution for $\delta\mathbf{u}_{k}$ is found, after $m$ iterations, by projecting on to the Krylov subspace
\begin{align*}
\mathcal{K}_{m}\left( \mathbf{J}_{k}\mathbf{P}^{-1},\mathbf{F}_{k} \right) &= \text{span}\left\{\mathbf{F}_{k},\mathbf{J}_{k}\mathbf{P}^{-1}\mathbf{F}_{k},\ldots,\left(\mathbf{J}_{k}\mathbf{P}^{-1}\right)^{m-1}\mathbf{F}_{k}\right\},
\end{align*}
where $\mathbf{J}_{k}=\mathbf{J}(\mathbf{u}_{k})$ and $\mathbf{F}_{k}=\mathbf{F}(\mathbf{u}_{k})$. The preconditioner matrix $\mathbf{P}\approx\mathbf{J}_{k}$ is a sparse approximation to the full Jacobian $\mathbf{J}_{k}$ and will be discussed further later. The purpose of the preconditioner is to reduce the number of iterations of the GMRES algorithm by reducing the dimension of the Krylov subspace necessary to find an accurate value for $\delta\mathbf{u}_{k}$.

The benefit of using a Krylov subspace method is that it does not require the full Jacobian to be formed. The Jacobian is only actioned on vectors to form the basis of the preconditioned Krylov subspace $\mathcal{K}_{m}$. First order quotients can be used to approximate these actions without having to explicitly form $\mathbf{J}_{k}$:
\begin{align*}
\mathbf{J}_{k}\mathbf{P}^{-1}\mathbf{v} &\approx \frac{\mathbf{F}\left(\mathbf{u}_{k}+h\mathbf{P}^{-1}\mathbf{v}\right) - \mathbf{F}(\mathbf{u}_{k})}{h},
\end{align*}
where $\mathbf{v}$ is an arbitrary vector used to form the Krylov subspace and $h$ is a sufficiently small shift \citep{Brown1990}.
Due to both the action of the Jacobian and the Newton correction both being approximated, we are left with an inexact Newton method, which leads to only superlinear and not quadratic convergence \citep{Knoll2004}. %Even though the rate of convergence is lower, the performance increase of the method makes up for the larger number of iterations.

The aim when forming the preconditioner matrix $\mathbf{P}$ is to construct an approximation to the Jacobian $\mathbf{J}_{k}$ that is easy to form and factorise, and exhibits a clustering of eigenvalues in the spectrum of the preconditioned Jacobian $\mathbf{J}_{k}\mathbf{P}^{-1}$. A good starting point for such a preconditioner is to consider a simplified form of the equations. In our case, we will be applying our integral formulation to the linearised problem that arises in the limit as the bottom topography approaches a flat bottom.

\section{The linear problem}\label{sec:linear}
\subsection{Exact solution}

In this section we consider the linearised version of (\ref{eq:new_finite_dynamic}) and (\ref{eq:new_finite_boundary_integral}) which arises from taking the limit that the perturbation of the bottom topography from a flat surface vanishes. This is done mathematically via a perturbation expansion in a parameter $\epsilon$ which is a measure of the ``height'' of the bottom disturbance.  For the particular example (\ref{eq:bottom}), the parameter $\epsilon$ is defined as the height of the Gaussian.

By taking the linear limit, the conditions on $z=\zeta(x,y)$ and $z=\beta(x,y)$ are projected onto the planes $z=0$ and $z=-1$, respectively.  Omitting the details, the linearised problem is to solve Laplace's equation
\begin{equation}
	\nabla^{2}\Phi = 0, \quad -1 < z < 0,
\label{eq:linearLaplace}
\end{equation}
subject to the kinematic, dynamic and radiation conditions \eqref{eq:new_finite_kinematic1}-\eqref{eq:new_finite_farfield1}, which are linearised under the perturbation expansion to become
\begin{align}
	\zeta_{x} &= \Phi_{z} & &z=0, \label{eq:linear_kinematic1} \\
	\beta_{x} &= \Phi_{z} & &z=-1, \label{eq:linear_kinematic2} \\
	\Phi_{x} - 1 + \frac{\zeta}{F^{2}} &= 0 & &z=0, \label{eq:linear_dynamic} \\
	(\Phi_{x},\Phi_{y},\Phi_{z}) &\rightarrow (1,0,0) & &x\rightarrow-\infty, \label{eq:linear_radiation1} \\
	\zeta &\rightarrow 0 & &x\rightarrow-\infty. \label{eq:linear_radiation2}
\end{align}

The linear problem (\ref{eq:linearLaplace})-(\ref{eq:linear_radiation2}) can be solved exactly using Fourier transforms to give
\begin{align}
\zeta(x,y) = &\frac{F^2}{2\pi^2}\int\limits_{-\pi/2}^{\pi/2}
\,\,\int\limits_{0}^{\infty}\frac{k^2\,\mathrm{sech}(k)\tilde{\beta}(k,\psi)\cos(k[|x|\cos\psi+y\sin\psi])}{kF^2-\sec^2\psi\tanh k}\,\,\mathrm{d}k\,\mathrm{d}\psi\notag\\
&-\frac{2 F^2 H(x)}{\pi}\int\limits_{\psi_0}^{\pi/2}\frac{k_1^2\,\mathrm{sech}(k_1)\tilde{\beta}(k_1,\psi) \sin(k_1x\cos\psi)\cos(k_1y\sin\psi)}{F^2-\sec^2\psi\,\mathrm{sech}^2k_1}\,\mathrm{d}\psi,\label{eq:exactsoln}
\end{align}
where $H(x)$ is the Heaviside function, $\tilde{\beta}(k,\psi)$ is the Fourier transform of $\beta(x,y)+1$ in polar coordinates, $\psi_0=0$ for $F<1$, $\psi_0=\arccos(1/F)$ for $F>1$ and the path of $k$-integration is taken below the pole $k=k_1$, where $k_1$ is the real positive root of $kF^2-\sec^2\psi\tanh k=0$. This formula \eqref{eq:exactsoln}  provides an extremely good approximation to the free-surface for flow regimes in which the bottom topography is almost flat.

It is interesting to note that the exact solution \eqref{eq:exactsoln} looks very similar to that for free-surface flow past a pressure distribution $p(x,y)$ that has been applied to the surface \citep{pethiyagoda15}.  In fact, the two problems are closely related.  Suppose we have a linear solution for flow past a bottom obstruction for a given bottom shape $z=\beta(x,y)$ and corresponding Fourier transform $\tilde{\beta}(k,\psi)$.  Then for the related linear problem of flow past a pressure distribution, if the pressure $p(x,y)$ is chosen so that its Fourier transform is given by
\begin{equation}
\tilde{p}(k,\psi)=\frac{\mathrm{sech}(k)\tilde{\beta}(k,\psi)}{\epsilon F^2},
\label{eq:pressure}
\end{equation}
then the wave pattern behind the pressure distribution is given by (\ref{eq:exactsoln}), except there is an addition term $-p(x,y)$ on the right-hand side.  That is, in the linear regime, the wave pattern downstream from the disturbance is the same.  Therefore, given that the problem of flow past an applied pressure is used as a model for studying ship wakes, there is a direct relationship between the wave patterns considered in this paper and ship waves.  We comment further on this relationship in the \S\ref{sec:discussion}.

\subsection{Boundary integral formulation}

Apart from benefiting from writing out the exact solution (\ref{eq:exactsoln}), the other reason for us to pursue the linearised problem is to construct a preconditioner for the fully nonlinear problem.  For this purpose we reformulate the linear problem (\ref{eq:linearLaplace})-(\ref{eq:linear_radiation2}) using Green's second theorem to give
\begin{align}
	2\pi\left(\phi^{*} - x^{*}\right) &= \iint_{\mathbb{R}^{2}} \frac{\zeta_{x}}{\sqrt{\left(x - x^{*}\right)^{2} + \left(y - y^{*}\right)^{2}}} \; \text{d}x\text{ d}y \nonumber \\
	& \quad {} + \iint_{\mathbb{R}^{2}} \frac{\beta_{x}}{\sqrt{ \left(x - x^{*}\right)^{2} + \left(y - y^{*}\right)^{2} + 1}} \; \text{d}x\text{ d}y \nonumber \\
	& \quad {} + \iint_{\mathbb{R}^{2}} \frac{(\psi - x^{*})}{\left( (x-x^{*})^{2} + (y-y^{*})^{2} + 1\right)^{3/2}} \; \text{d}x\text{ d}y,
\label{eq:linear_integral}
\end{align}
which is coupled to the dynamic condition
\begin{align}
\phi_{x}^{*} + \frac{\zeta^{*}}{F^{2}} - 1 &= 0. \label{eq:linear_dynamic2}
\end{align}
By applying the same discretisation to that described in \S\ref{sec:numerical_collocation}, we can derive a completely analogous system of $3(N+1)M$ equations for the same $3(N+1)M$ unknowns (\ref{eq:unknowns}).  The resulting equations in their exact form are detailed in Appendix~\ref{sec:linear_jacobian}.  We are interested in exploiting the structure and entries of the Jacobian, $\tilde{\mathbf{J}}$, of this new linear system.

\begin{figure}
\centering
\begin{subfigure}[t]{0.45\linewidth}
\centering
\includegraphics[width=0.9\linewidth]{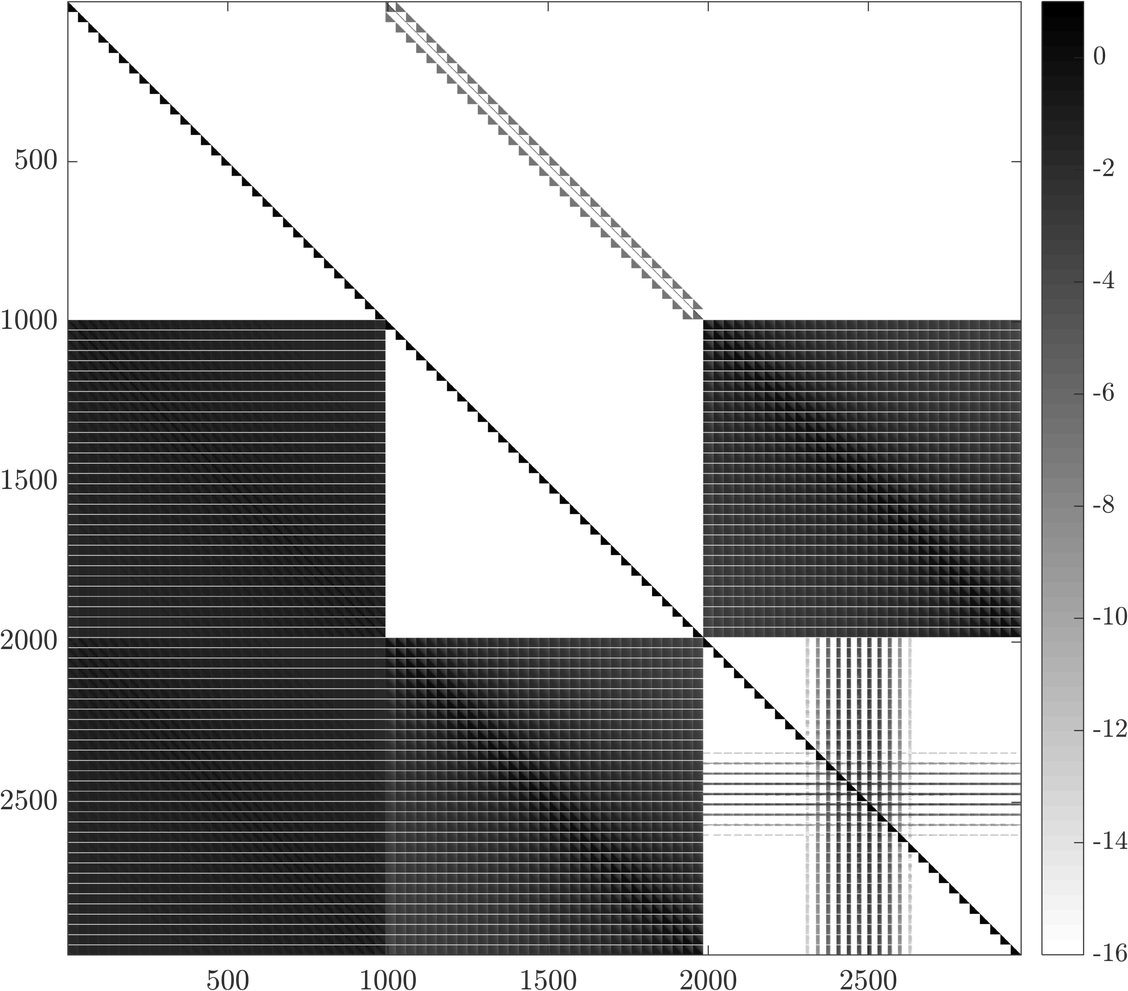}
\caption{Nonlinear Jacobian}
\label{fig:nonlinear_jac}
\end{subfigure}
\begin{subfigure}[t]{0.45\linewidth}
\centering
\includegraphics[width=0.9\linewidth]{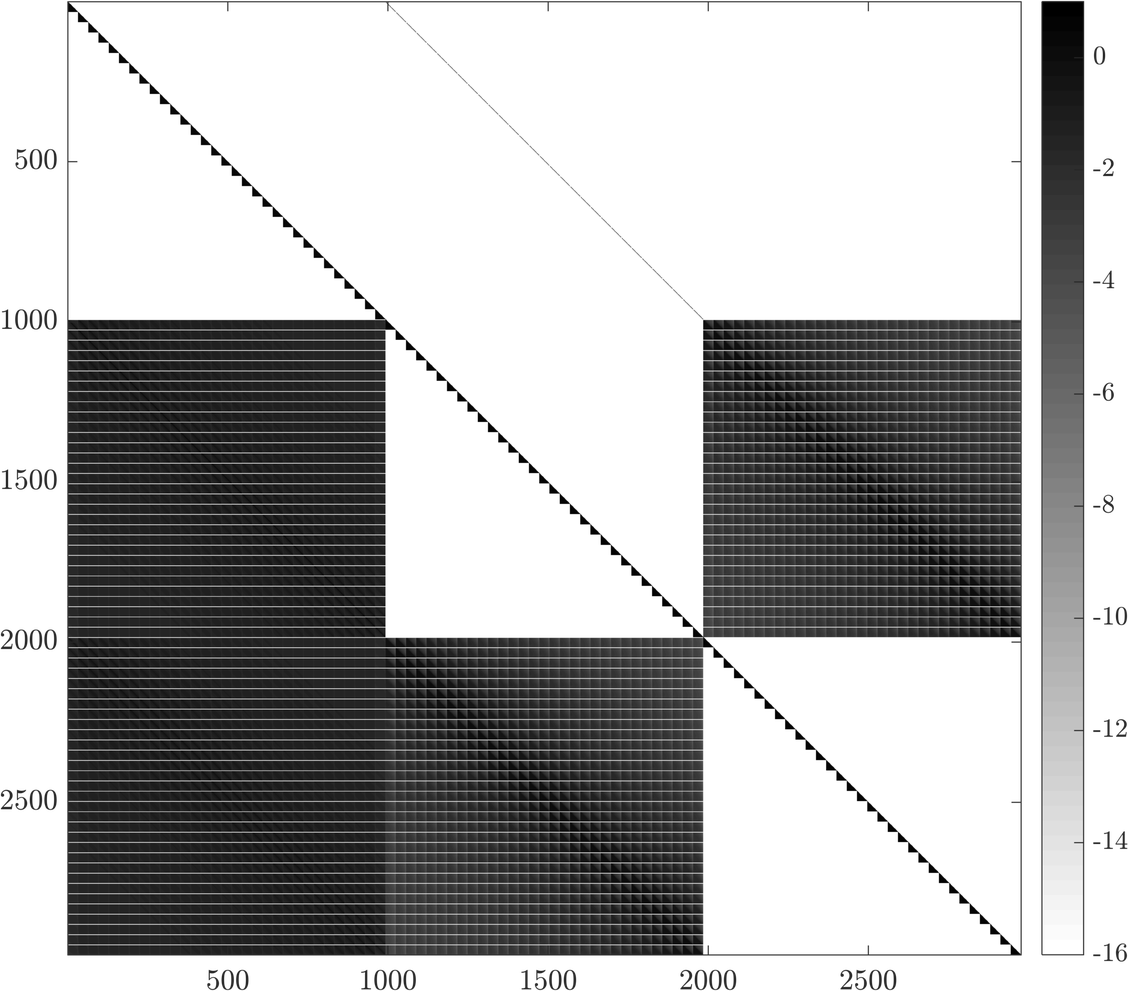}
\caption{Linear Jacobian}
\label{fig:linear_jac}
\end{subfigure}
\caption{Visualisation of the nonlinear and linear Jacobians showing the logarithm (base 10) of the magnitudes of the entries. Entries with larger magnitudes are represented by darker shades { and zero entries are coloured white}. Jacobians were generated from the problem on a $31 \times 31$ mesh with $\Delta x = \Delta y = 0.8$, $F=0.6$, a single Gaussian bump on the bottom surface with scale parameter $\epsilon = 0.1$. Figure 2(a) shows the full Jacobian for the completely nonlinear problem, which is generated using finite differences, while 2(b) shows the Jacobian of the linearised problem. The linear Jacobian can be generated analytically by taking derivatives of the linearised equations. The two Jacobians clearly share a similar structure, with dense blocks and sparse blocks in the same places. The key differences are the top-middle block having more of a diagonal structure and the bottom right block only having the block diagonal structure in the linear case. The structural similarities suggest that the linear Jacobian may be a good approximation for the nonlinear Jacobian.}
\label{fig:jacobian_compare}
\end{figure}

Figure \ref{fig:jacobian_compare} shows the similarity between the nonlinear Jacobian $\mathbf{J}_{k}$ and the Jacobian of the corresponding linear problem $\tilde{\mathbf{J}}$. The visualisation shows that $\tilde{\mathbf{J}}$ shares the structure of $\mathbf{J}_{k}$ while being much easier to form as analytical formulas exist for the derivatives which are provided in their exact form in Appendix~\ref{sec:linear_jacobian}. Since analytical formulas exist for the elements of $\tilde{\mathbf{J}}$, the elements can be calculated independently and utilise parallel processing. We are also able to exploit the block structure of the matrix by forming the blocks separately. This is useful as $\tilde{\mathbf{J}}_{2,3} = \tilde{\mathbf{J}}_{3,2}$, where $\tilde{\mathbf{J}}$ is treated as a $3\times3$ block matrix, and $\tilde{\mathbf{J}}_{2,2} = \tilde{\mathbf{J}}_{3,3}$, which allows us to save time by not explicitly forming two blocks.

\begin{figure}
	\centering
	\begin{subfigure}[t]{0.45\linewidth}
		\centering
		\includegraphics[width=\linewidth]{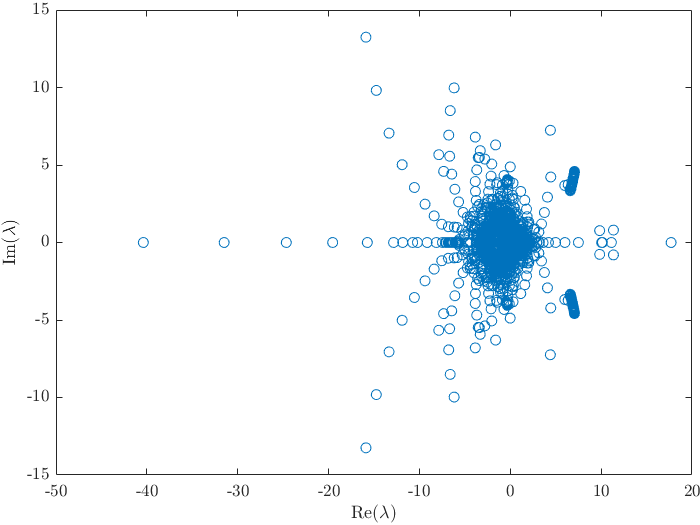}
		\caption{Nonlinear Jacobian $\mathbf{J}_{k}$ eigenvalues}
		\label{fig:full_eigenvalues1}
	\end{subfigure}
	\begin{subfigure}[t]{0.45\linewidth}
		\centering
		\includegraphics[width=\linewidth]{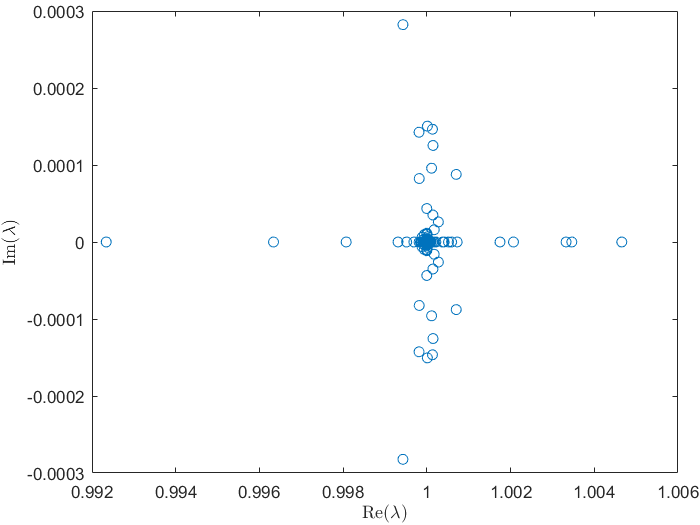}
		\caption{Preconditioned Jacobian $\mathbf{J}_{k}\tilde{\mathbf{J}}^{-1}$ eigenvalues}
		\label{fig:full_eigenvalues2}
	\end{subfigure}
	\\
	\begin{subfigure}[t]{0.45\linewidth}
		\centering
		\includegraphics[width=\linewidth]{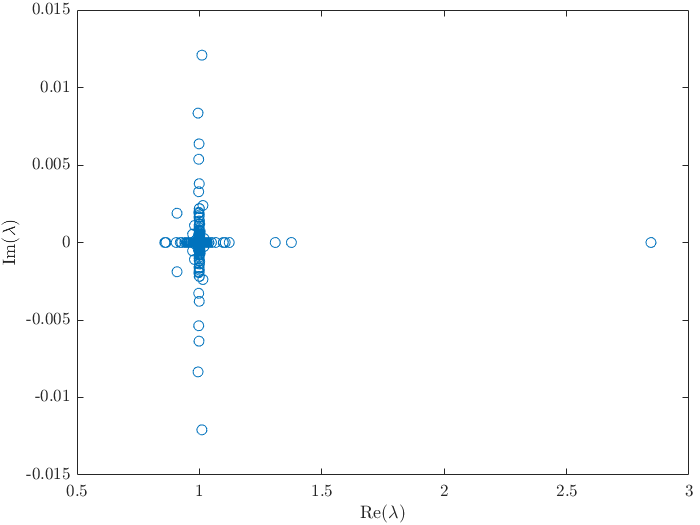}
		\caption{Full Jacobian preconditioned with banded linear Jacobian $\mathbf{J}_{k}\mathbf{P}^{-1}$}
		\label{fig:full_eigenvalues3}
	\end{subfigure}
	\caption{Eigenvalue plots of (a) the full nonlinear Jacobian $\mathbf{J}_{k}$ for an initial guess of a flat surface with $F=0.6$ and $\epsilon=0.1$ on a $31 \times 31$ mesh with $\Delta x = \Delta y = 0.8$ and (b) the full nonlinear Jacobian preconditioned with the linear Jacobian, $\mathbf{J}_{k}\tilde{\mathbf{J}}^{-1}$. The effect of preconditioning is a tight clustering of the eigenvalues, $\lambda$, around $1$. This tight clustering is desirable as a smaller spectrum of the preconditioned Jacobian, $\sigma(\mathbf{J}_{k}\tilde{\mathbf{J}}^{-1})$, will result in a smaller Krylov subspace required to find a solution to the Newton step $\delta\mathbf{u}_{k}$.}
	\label{fig:full_eigenvalues}
\end{figure}

An effective preconditioner has to reduce the dimension of the Krylov subspace required to find a solution. To test whether the linear Jacobian is an effective preconditioner, we can compare the eigenvalues, $\lambda$,  of the nonlinear Jacobian and the preconditioned nonlinear Jacobian. Figure \ref{fig:full_eigenvalues}(a) shows the eigenvalues of the nonlinear Jacobian $\mathbf{J}_{k}$ for a flat surface. Figure \ref{fig:full_eigenvalues}(b) shows the eigenvalues for the nonlinear Jacobian preconditioned with the linear Jacobian, that is $\mathbf{J}_{k}\tilde{\mathbf{J}}^{-1}$. The clustering of eigenvalues is much tighter than the non preconditioned Jacobian. Thus, we chose that $\mathbf{P}=\tilde{\mathbf{J}}$.

Now that we have a preconditioner $\mathbf{P}$ that is cheap to form, we want to find a way to reduce the large memory requirement to store $\mathbf{P}$. In the four dense blocks in $\mathbf{P}$ the magnitude of the elements decays to zero away from the main diagonal. This structure suggests that it is possible to store only the elements near the diagonal using a banded approximation of the dense blocks. Indeed, we follow this approach and use a banded version of $\tilde{\mathbf{J}}$ for our preconditioner $\mathbf{P}$. The details are described in Appendix~\ref{app:banded}. Figure \ref{fig:full_eigenvalues}(c) shows the eigenvalues for the nonlinear Jacobian preconditioned with the banded preconditioner, $\mathbf{J}_{k}\mathbf{P}$. While the clustering is not as tight as when using the full linear Jacobian, the dimension of the Krylov subspace required to find a solution is still significantly reduced.

\section{Numerical results}\label{sec:results}
Results were computed using MATLAB on either a standard desktop computer with a maximum of $16$ GB of system memory or a high performance computer with up to $450$ GB of system memory.  Apart from system memory, the high performance computer also benefits from a higher number of processing cores, which allows parallelised code to perform faster.  In all cases, we utilised the KINSOL implementation of the Jacobian-free Newton-Krylov method \citep{Hindmarsh2005}.  For the following, note that an $N\times M$ mesh means $N$ grid points in the $x$ direction and $M$ grid points in the $y$ direction.

\subsection{Subcritical flow}
\begin{figure*}
\centering
\begin{subfigure}[t]{0.6\linewidth}
\centering
\raisebox{1em}{\includegraphics[width=\linewidth]{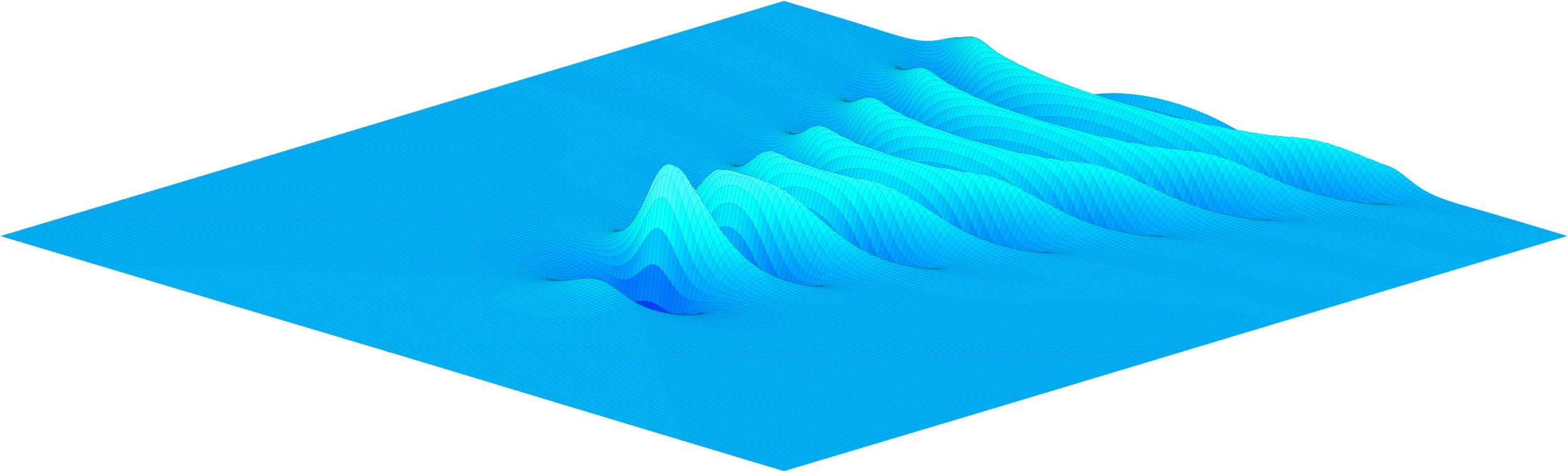}}
\caption{Linear solution}
\label{fig:subcritical1}
\end{subfigure}
\begin{subfigure}[t]{0.3\linewidth}
\centering
\includegraphics[width=\linewidth]{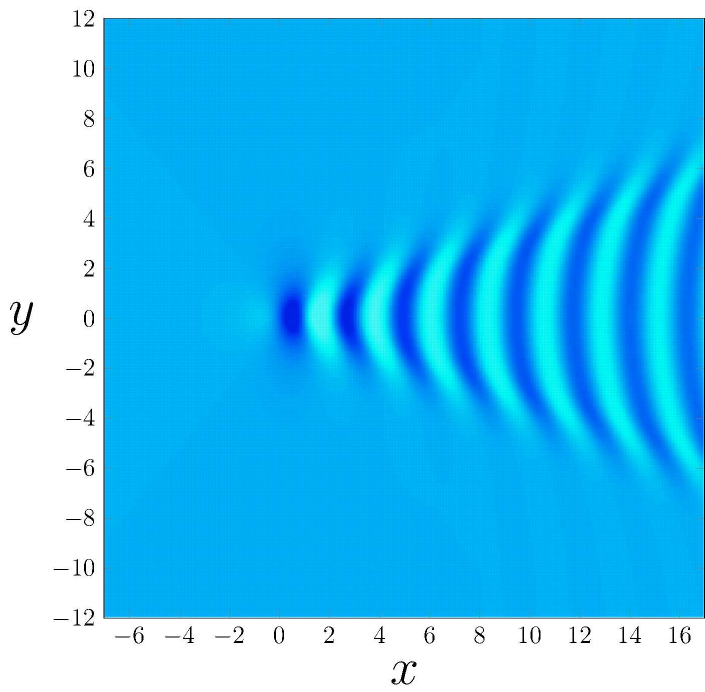}
\caption{Plan view}
\label{fig:subcritical2}
\end{subfigure}
\\ \vspace{1em}
\begin{subfigure}[t]{0.6\linewidth}
\centering
\raisebox{1em}{\includegraphics[width=\linewidth]{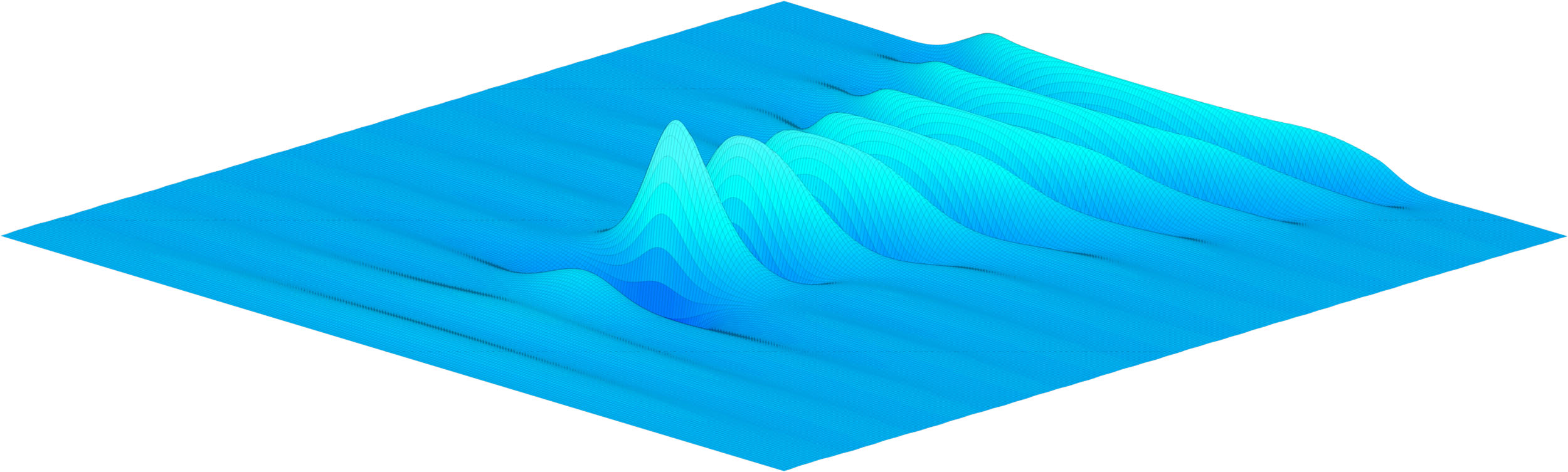}}
\caption{Nonlinear solution, $\epsilon=0.1$}
\label{fig:subcritical3}
\end{subfigure}
\begin{subfigure}[t]{0.3\linewidth}
\centering
\includegraphics[width=\linewidth]{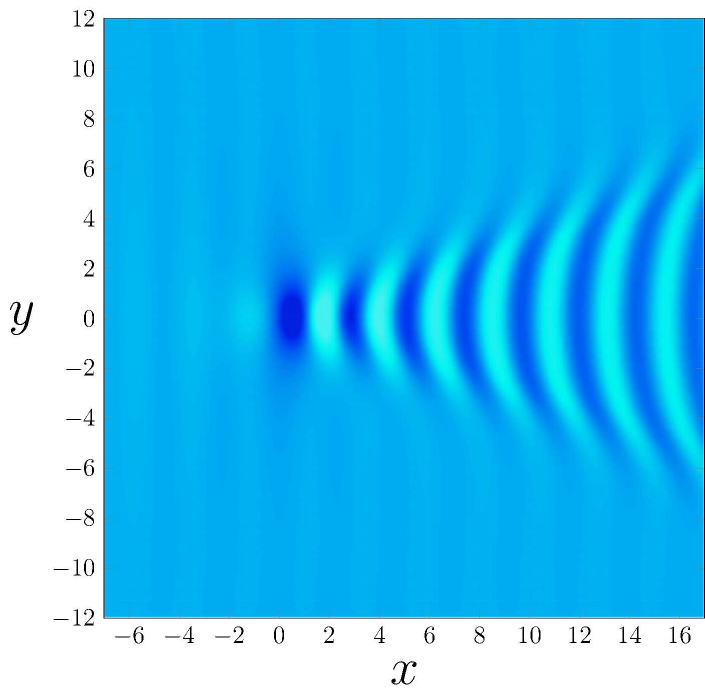}
\caption{Plan view}
\label{fig:subcritical4}
\end{subfigure}
\\ \vspace{1em}
\begin{subfigure}[t]{0.6\linewidth}
\centering
\raisebox{1em}{\includegraphics[width=\linewidth]{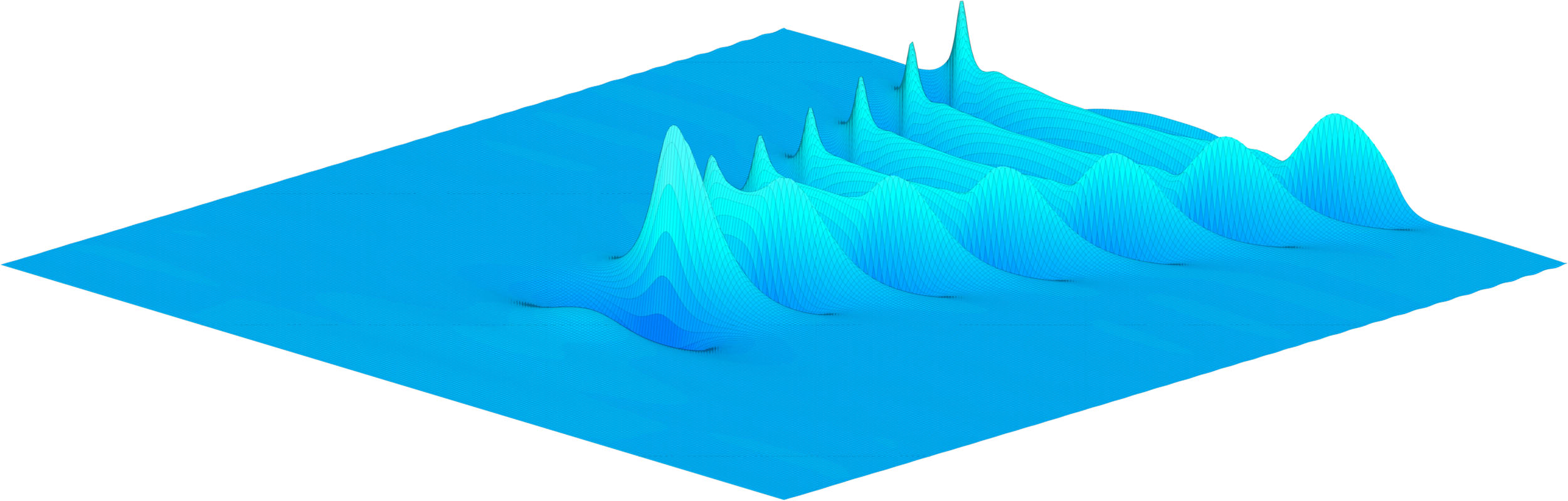}}
\caption{Nonlinear solution, $\epsilon=0.285$}
\label{fig:subcritical5}
\end{subfigure}
\begin{subfigure}[t]{0.3\linewidth}
\centering
\includegraphics[width=\linewidth]{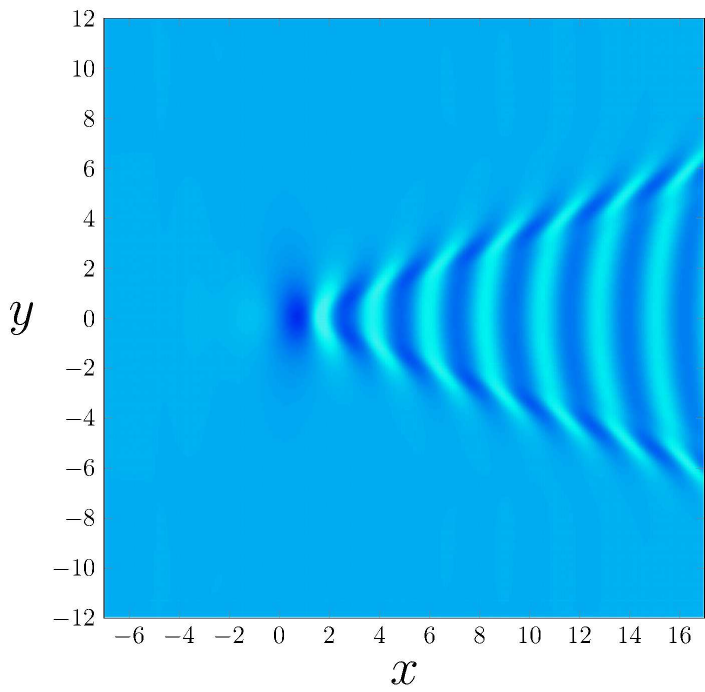}
\caption{Plan view}
\label{fig:subcritical6}
\end{subfigure}
\\ \vspace{1em}
\begin{subfigure}[t]{0.7\linewidth}
	\centering
	\includegraphics[width=\linewidth]{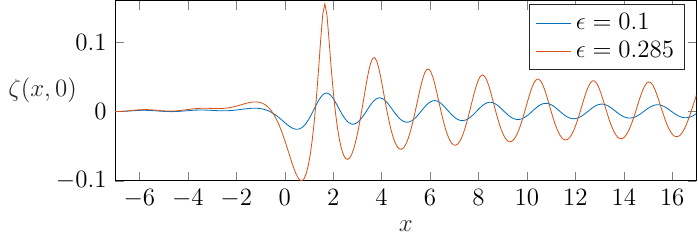}
	\caption{{ Centreline plots for $\epsilon=0.1$ and $0.285$.}}
	\label{fig:subcritical7}
\end{subfigure}
\caption{Numerical solutions for subcritical flow past a single bump (\ref{eq:bottom}) centred at the origin. Solutions were computed for $-7\leq x\leq 17$, $-12\leq y\leq 12$ using a $281\times281$ mesh with $\Delta x = \Delta y = 0.0857$, $\delta=0.5$ and $F = 0.6$. (a)-(b) show the linear solution, (c)-(d) show the nonlinear solution for $\epsilon=0.1$ and (e)-(f) show the nonlinear solution for $\epsilon=0.285$. {  (g) presents a comparison of the two nonlinear surfaces along the centreline $y=0$.}}
\label{fig:subcritical}
\end{figure*}

Subcritical flows are defined by $F<1$.  In this regime, the wave pattern is characterised by the presence of both divergent and transverse waves.  In Figure~\ref{fig:subcritical}, we provide some representative results for a fixed subcritical Froude number $F=0.6$.  In all cases, the bottom topography takes the form (\ref{eq:bottom}) with $\delta=0.5$, which represents a single Gaussian-type bump with height $\epsilon$ on an otherwise flat bottom.

Figure~\ref{fig:subcritical}(a)-(b) shows the wave pattern for the exact linear solution (\ref{eq:exactsoln}), which is valid for small bump heights, $\epsilon\ll 1$.  We can see that, for this linear solution, the wave pattern appears to be dominated by the transverse waves which run roughly perpendicular to the direction of flow.  Similar free-surface profiles are provided in Figure~\ref{fig:subcritical}(c)-(d).  This time the bump height was fixed to be $\epsilon=0.1$ and the solution was computed using our fully nonlinear numerical scheme.  Thus we can see for small ($\epsilon\ll 1$) to moderately small ($\epsilon=0.1$) obstructions, the linear and nonlinear results are not significantly different.

For larger values of $\epsilon$, differences emerge between the linear and nonlinear regimes, and these differences become more pronounced as $\epsilon$ increases.  For example, in Figure~\ref{fig:subcritical}(e)-(f) we present results for $\epsilon=0.285$, which represents a highly nonlinear solution.  In Figure~\ref{fig:subcritical}(e) in particular, we see a remarkably nonlinear wave pattern with sharp divergent waves shaped like dorsal fins, whose amplitude is much higher than the transverse waves (these are reminiscent of the highly nonlinear solutions computed by \citet{pethiyagoda2014} for flow past a submerged source singularity).  It is important to emphasise that { the distinctive shape of the sharp divergent waves} are not predicted by linear theory, demonstrating that for sufficiently large disturbances to the free stream, a nonlinear formulation is required to accurately describe the wave pattern.

{ As an indication of the relative vertical scales involved, we present in Figure~\ref{fig:subcritical}(g) a plot of the two nonlinear surfaces along the centreline $y=0$.  Clearly the amplitude of the waves for $\epsilon=0.285$ is significantly higher than that for $\epsilon=0.1$, as expected.  Further, the waves for $\epsilon=0.285$ appear to have broader troughs and sharper creasts.  Another feature of Figure~\ref{fig:subcritical}(g) is that even though the Froude number is fixed, the wavelength is smaller for the more nonlinear solution.  These are all commonly observed properties of nonlinear free surface flows in two dimensions (\cite{Forbes1982,Forbes1985,mccue2002}).}

We note that for $F=0.6$, we did not compute solutions for $\epsilon>0.285$.  From a mathematical perspective, we expect that solutions will exist up to some maximum value of $\epsilon$ at which point the surface will be in some limiting configuration which is likely to have the maximum surface height at $F^2/2$ (Bernoulli's equation (\ref{eq:finite_dynamic}) does not allow for the free-surface to be higher than $F^2/2$).  This limiting configuration would be some type of complicated three-dimensional analogue of Stokes limiting configuration in two dimensions (with a $120^\circ$ corner at the crest).

{ Returning to the free-surface profile in Figure~\ref{fig:subcritical}(c)-(d), small amplitude waves appear ahead of the bottom disturbance which are not physical and are caused by truncating the domain upstream.  These spurious numerical artifacts appear in all of our solutions for subcritical Froude numbers, and can be moderated by truncating further upstream (which, for a fixed grid spacing, involves increasing the number of grid points) or by varying the parameter $n$ in the radiation condition (\ref{eq:scullen1})-(\ref{eq:scullen6}).   Note that these numerical waves appear larger in Figure~\ref{fig:subcritical}(c)-(d) than in Figure~\ref{fig:subcritical}(e)-(f), however in fact they are similar in size (see Figure~\ref{fig:subcritical}(g), for example).  It is the vertical scaling of the surfaces that causes this illusion.}

\subsection{Supercritical flow}
\begin{figure*}
\centering
\begin{subfigure}[t]{0.6\linewidth}
\centering
\raisebox{2em}{\includegraphics[width=\linewidth]{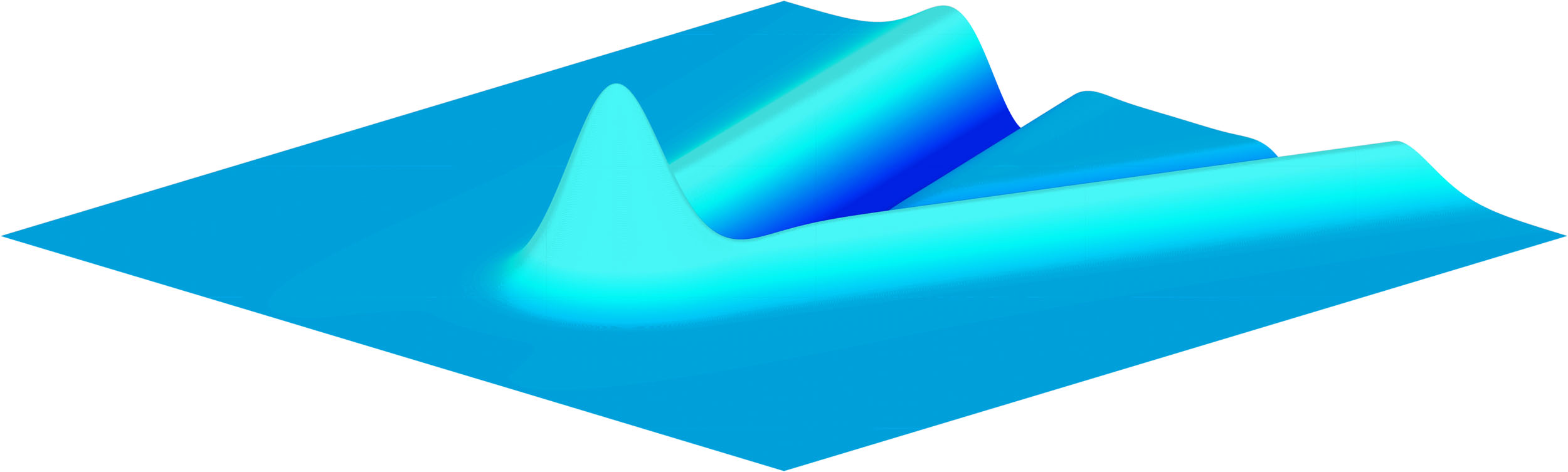}}
\caption{Linear solution}
\label{fig:supercritical1}
\end{subfigure}
\begin{subfigure}[t]{0.3\linewidth}
\centering
\includegraphics[width=\linewidth]{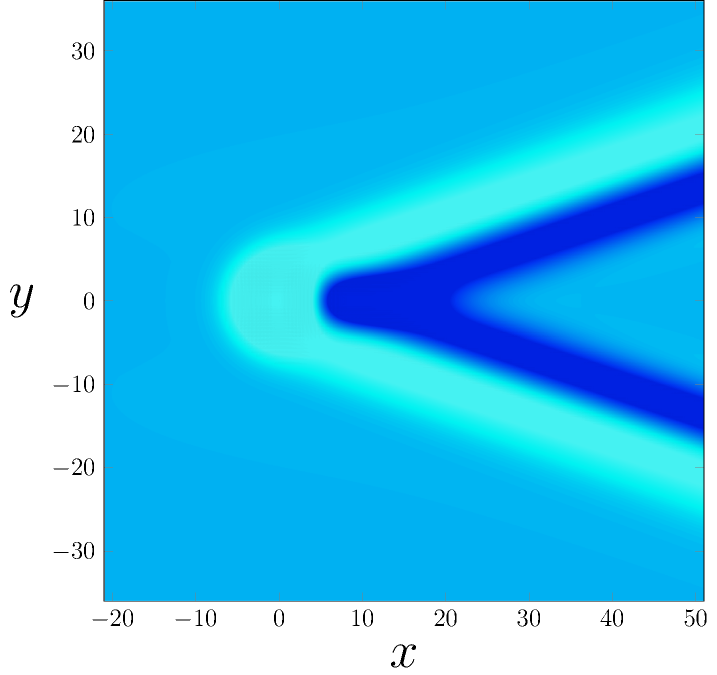}
\caption{Plan view}
\label{fig:supercritical2}
\end{subfigure}
\\ \vspace{1em}
\begin{subfigure}[t]{0.6\linewidth}
\centering
\raisebox{2em}{\includegraphics[width=\linewidth]{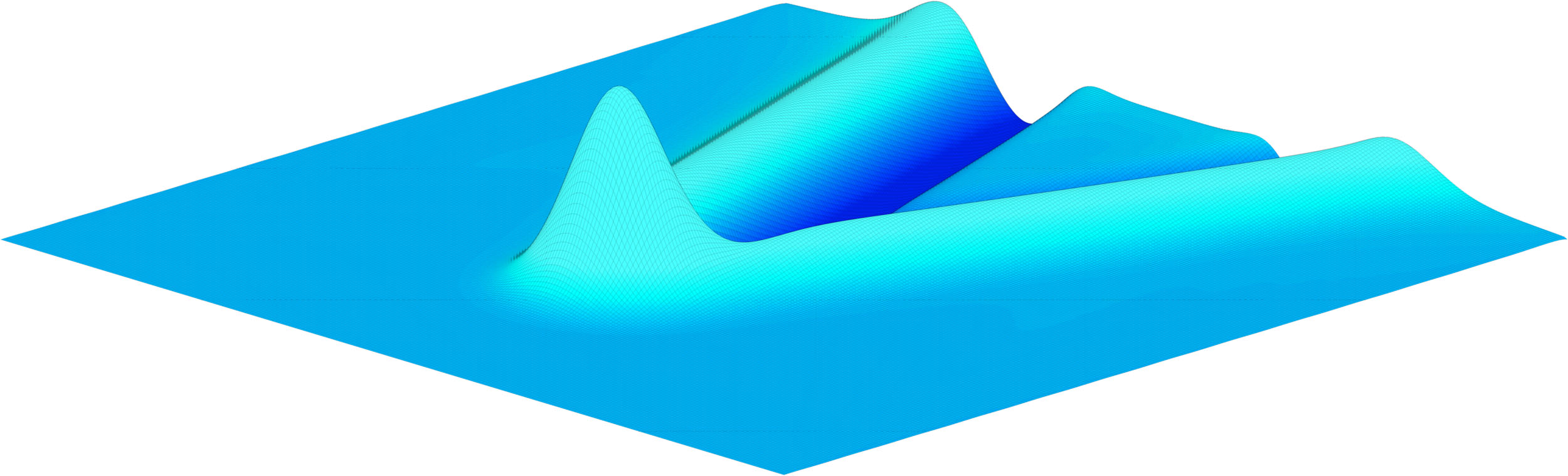}}
\caption{Nonlinear solution, $\epsilon=0.1$}
\label{fig:supercritical3}
\end{subfigure}
\begin{subfigure}[t]{0.3\linewidth}
\centering
\includegraphics[width=\linewidth]{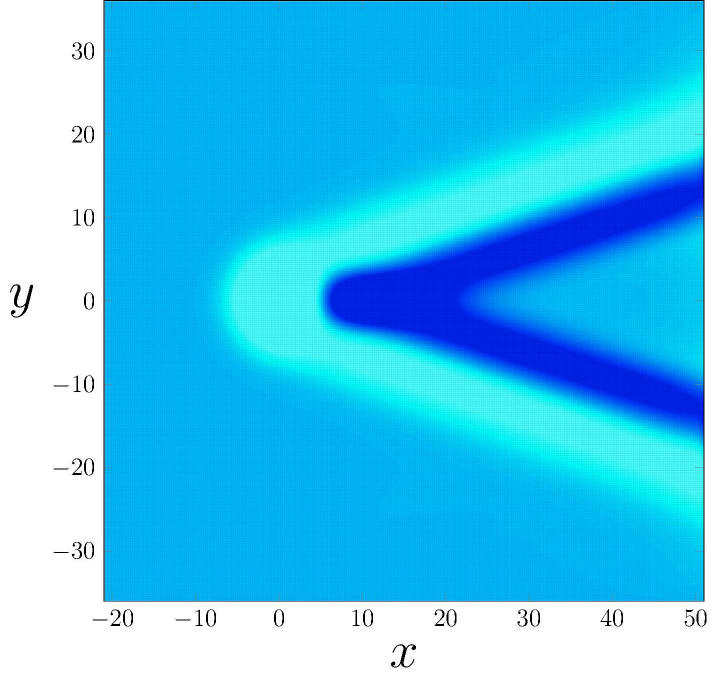}
\caption{Plan view}
\label{fig:supercritical4}
\end{subfigure}
\\ \vspace{1em}
\begin{subfigure}[t]{0.6\linewidth}
\centering
\raisebox{2em}{\includegraphics[width=\linewidth]{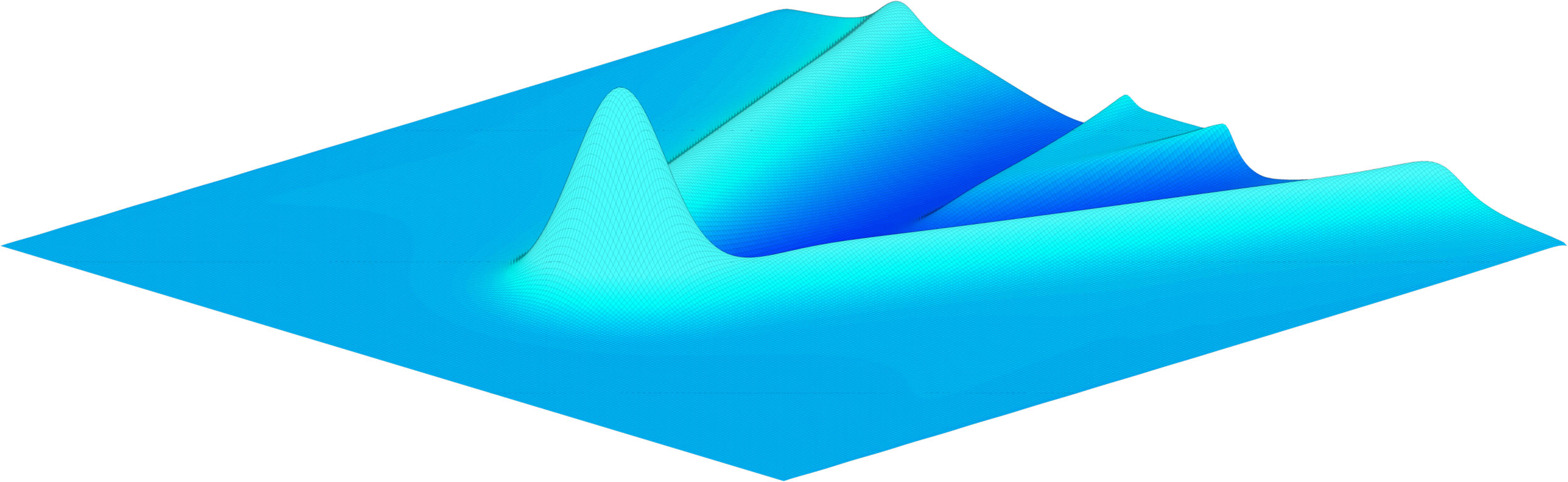}}
\caption{Nonlinear solution, $\epsilon=2.75$}
\label{fig:supercritical5}
\end{subfigure}
\begin{subfigure}[t]{0.3\linewidth}
\centering
\includegraphics[width=\linewidth]{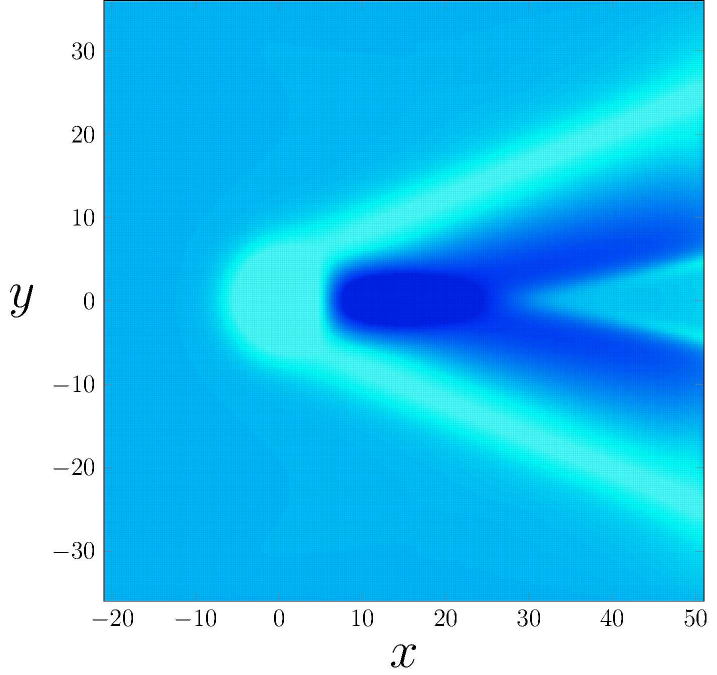}
\caption{Plan view}
\label{fig:supercritical6}
\end{subfigure}
\\ \vspace{1em}
\begin{subfigure}[t]{0.7\linewidth}
\centering
\includegraphics[width=\linewidth]{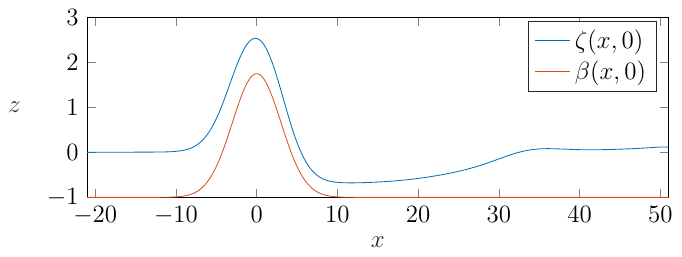}
\caption{{ Centreline plots for $\epsilon=2.75$}}
\label{fig:supercritical7}
\end{subfigure}
\caption{Numerical solutions for supercritical flow past a single bump (\ref{eq:bottom}) centred at the origin. Solutions were computed for $-21\leq x\leq 51$, $-36\leq y\leq 36$ using a $281\times281$ mesh with $\Delta x = \Delta y = 0.257$, $x_{1}=-21$, $\delta=3$ and $F = 3$. (a)-(b) show the linear solution, (c)-(d) show the nonlinear solution with $\epsilon=0.1$ and (e)-(f) show the nonlinear solution with $\epsilon=2.75$.
As an indication of the relative vertical scales in these plots, the maximum height for the solution in (c)-(d) is $\zeta(x,0)=1.091$ while the maximum height for the solution in (e)-(f) is $\zeta(x,0)=2.556$. { (g) shows a cut along the centreline for the solution with $\epsilon=2.75$, including the free surface $\zeta(x,0)$ and the bottom $\beta(x,0)$.}}
\label{fig:supercritical}
\end{figure*}

Supercritical flows are for $F>1$.  In Figure~\ref{fig:supercritical}, we present supercritical solutions for the representative value $F=3$.  Again, we have chosen the bottom obstruction to be a single bump of the form (\ref{eq:bottom}), but now with $\delta=3$.  Figure~\ref{fig:supercritical}(a)-(b) shows free-surface profiles for the exact linear solution (\ref{eq:exactsoln}), while Figure~\ref{fig:supercritical}(c)-(d) shows the surfaces for the bump height $\epsilon=0.1$.  The first point to note here is that the wave pattern appears very different to that presented for subcritical flows.  This qualitative difference is well known, { and is for the most part due to supercritical wave patterns being entirely made up of divergent waves.  According to linear theory, the mathematical reason for the absence of transverse waves is that the dispersion relation $kF^2-\sec^2\psi\tanh k=0$ has no real positive roots for $0\leq \psi < \arccos(1/F)$.  Physically, if we consider an extremely localised disturbance, then for supercritical flow the surface elevation at a single point downstream is due to a wave that began propagating out from the disturbance at the group velocity some time ago, while for subcritical flow the elevation at a single point is due to two waves that began propagating at two different previous times (geometric constructions like these are explained in detail in \cite{Lamb1932,pethiyagoda17b,soomere2007,wehausen60}, for example.)}  The second point to note is that for such a small value as $\epsilon=0.1$, the nonlinear wave pattern appears very similar to the linear solution.

We are motivated to increase the bump height $\epsilon$ to observe the effects of nonlinearity.  It turns out that we are able to compute numerical solutions to the fully nonlinear problem for much larger values of $\epsilon$.  For example, in Figure~\ref{fig:supercritical}(e)-(f) we show free-surface profiles for $\epsilon=2.75$.  Given that $\epsilon$ measures the height of the Gaussian (\ref{eq:bottom}), this value of $\epsilon$ corresponds to a significant deviation from a flat bottom.  { To illustrate the scales involved, we have presented in Figure~\ref{fig:supercritical}(g) a slice of the solution along the centreline $y=0$, including the free surface $z=\zeta(x,0)$ and the bottom boundary $z=\beta(x,0)$.}  This type of solution is very interesting because the bottom disturbance is much higher than the depth of the channel far upstream, even though the bump itself does not pierce the surface.

Further, we can see some features in Figure~\ref{fig:supercritical}(e)-(f) that are not present in the linear solution.  For example: the V-shaped wave ridge is much steeper for $\epsilon=2.75$ than the linear solution; the angle of this V-shaped ridge (the apparent wake angle) is larger than that for the linear solution; and there appears to be a second, smaller, V-shaped wedge that forms a distance behind the disturbance.  This is a new feature that is not able to be predicted by linear theory.

\begin{figure*}
	\centering
	\begin{subfigure}[t]{0.48\textwidth}
		\centering
		\includegraphics[width=0.9\linewidth]{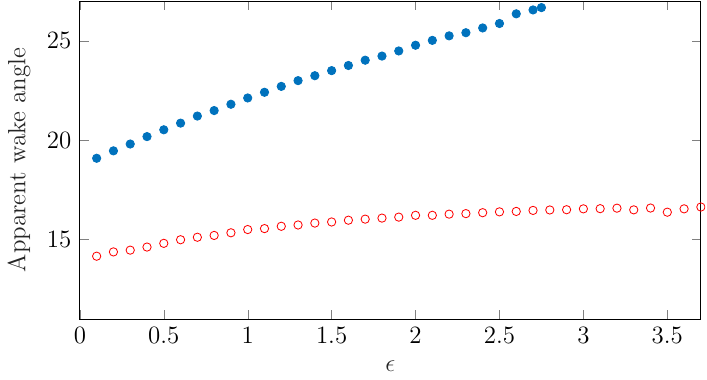}%
	\end{subfigure}
	\begin{subfigure}[t]{0.48\textwidth}
		\centering
		\includegraphics[width=0.9\linewidth]{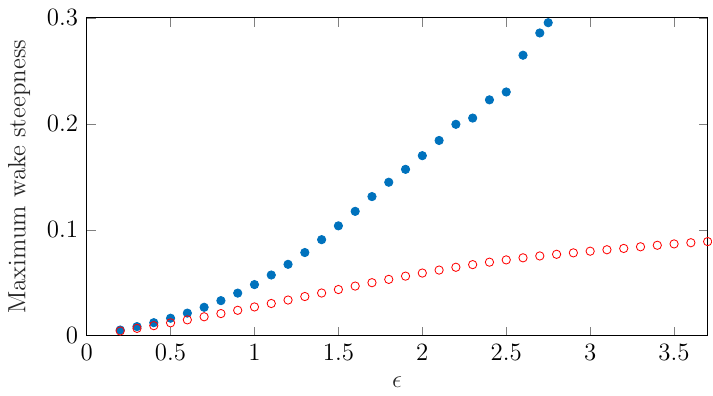}
	\end{subfigure}
	\caption{{ Plots of the (a) apparent wake angle and (b) wave steepness against bump height $\epsilon$ for $F = 3$ (solid blue circles) and $4$ (open red circles).  The apparent wake angle is given by the angle between the V-shaped ridge and the centreline.  The wave steepness is defined as the maximum slope of the surface on the outside of the V-shaped ridge.  These results demonstrate that nonlinearity has a significant effect on the wave pattern for supercritical flow regimes.}}
	\label{fig:wake_angle_fr_3}
\end{figure*}

{ To investigate these nonlinear effects further, we show in Figure~\ref{fig:wake_angle_fr_3} the dependence of the apparent wake angle on the obstruction height $\epsilon$ for two supercritical Froude numbers, $F=3$ and $4$.  These two plots show how the wake angle is significantly lower for the larger Froude number, which is true for both linear and nonlinear regimes.  In addition, the increase in wake angle with nonlinearity is clear from these plots.  Also shown in Figure~\ref{fig:wake_angle_fr_3} is a plot of the maximum steepness of the surface versus the obstruction height $\epsilon$.  Here wave steepness is defined to be the maximum slope on the outside of the (main) V-shaped ridge that extends to the far-field.  We see that the wave steepness clearly increases with nonlinearity.  Presumably this steepness will continue to increase as $\epsilon$ increases until the surface near the origin reaches its limiting height of $F^2/2$.}

\subsection{Multiple bumps}
\begin{figure*}
\centering
\begin{subfigure}[t]{0.6\linewidth}
\centering
\raisebox{2em}{\includegraphics[width=\linewidth]{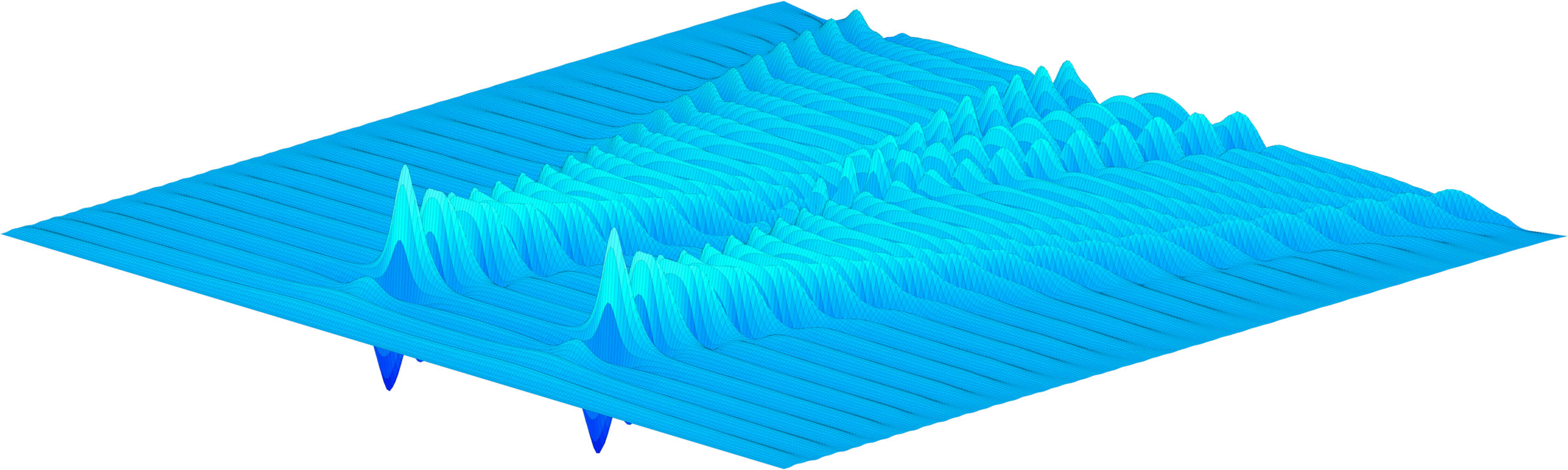}}
\caption{Symmetrical bumps}
\label{fig:multibumps1}
\end{subfigure}
\begin{subfigure}[t]{0.3\linewidth}
\centering
\includegraphics[width=\linewidth]{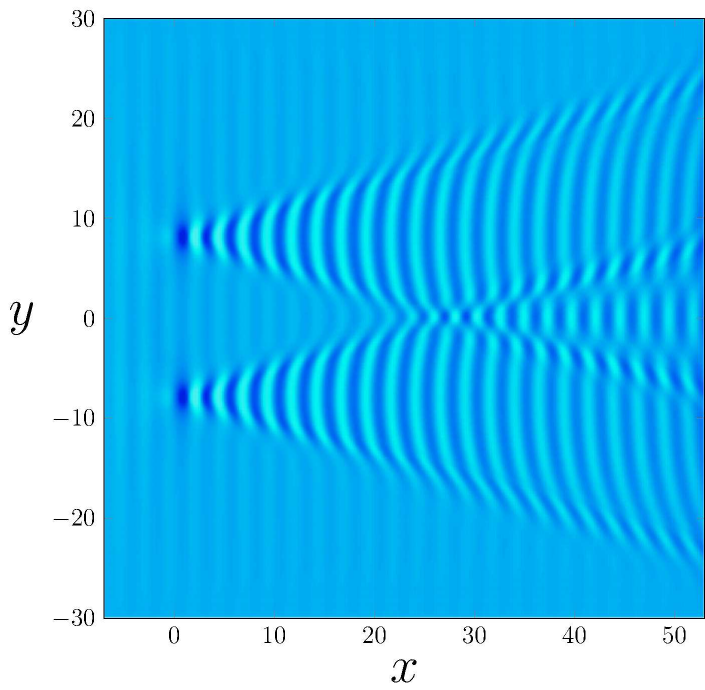}
\caption{Plan view}
\label{fig:multibumps2}
\end{subfigure}
\\ \vspace{1em}
\begin{subfigure}[t]{0.6\linewidth}
\centering
\raisebox{2em}{\includegraphics[width=\linewidth]{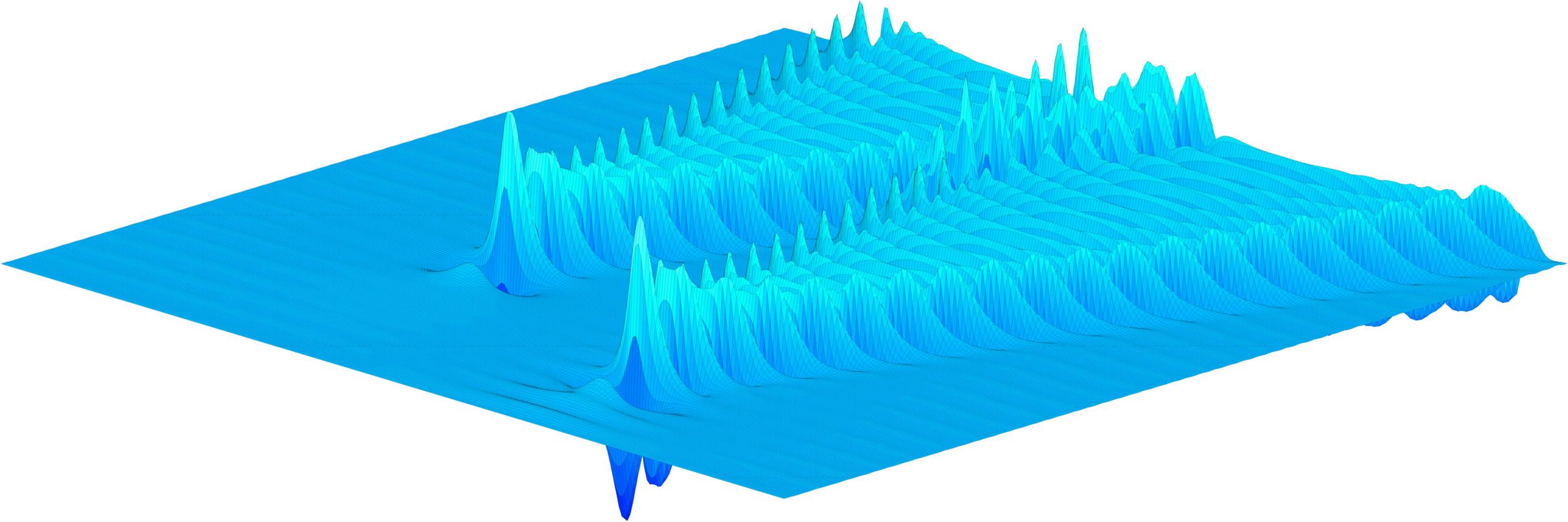}}
\caption{Asymmetrical bumps}
\label{fig:multibumps3}
\end{subfigure}
\begin{subfigure}[t]{0.3\linewidth}
\centering
\includegraphics[width=\linewidth]{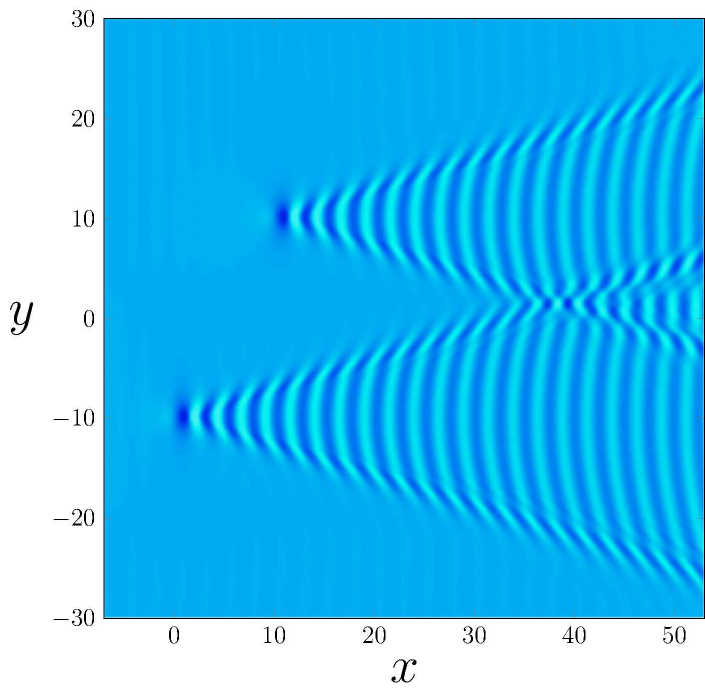}
\caption{Plan view}
\label{fig:multibumps4}
\end{subfigure}
\caption{Numerical solutions for subcritical flow past two bumps of the form (\ref{eq:bottom}). Solutions were computed for $-7\leq x\leq 53$, $-30\leq y\leq 30$  using a $281\times281$ mesh with $\Delta x = \Delta y = 0.2143$, $F= 0.6$, bump height $\epsilon=0.2$, $\delta=0.5$ and the upstream truncation point $x_{1}=-7$. (a)-(b) show the solution for flow past two bumps positioned symmetrically across the $x$-axis at $(0,8)$ and $(0,-8)$, (c)-(d) show the solution for flow past two bumps offset in the $x$ direction at $(10,10)$ and $(0,-10)$.}
\label{fig:multibumps}
\end{figure*}

An advantage of formulating our numerical scheme with an arbitrarily defined bottom topography $z=\beta(x,y)$ is that solutions can easily be computed for flow over multiple bumps on the bottom surface.  Two such flow configurations are shown in Figure~\ref{fig:multibumps} for a fixed Froude number $F=0.6$.  Figure \ref{fig:multibumps}(a)-(b) shows the numerical solution to the nonlinear problem with two bumps positioned symmetrically across the $x$-axis. The wave trains appear independent for sufficiently small $x$ and then interfere with each other at roughly $x=30$. In Figure~\ref{fig:multibumps}(c)-(d) we show the wave pattern for flow past two bumps positioned asymmetrically across the $x$-axis.  In this configuration, one of the bumps is further downstream than the other one. Again, the wave trains interfere with each other, as expected.  Note for this configuration there is no direct analogue in two dimensions.

\begin{figure*}
	\centering
	\begin{subfigure}[t]{0.6\textwidth}
		\centering
		\raisebox{2em}{\includegraphics[width=\textwidth]{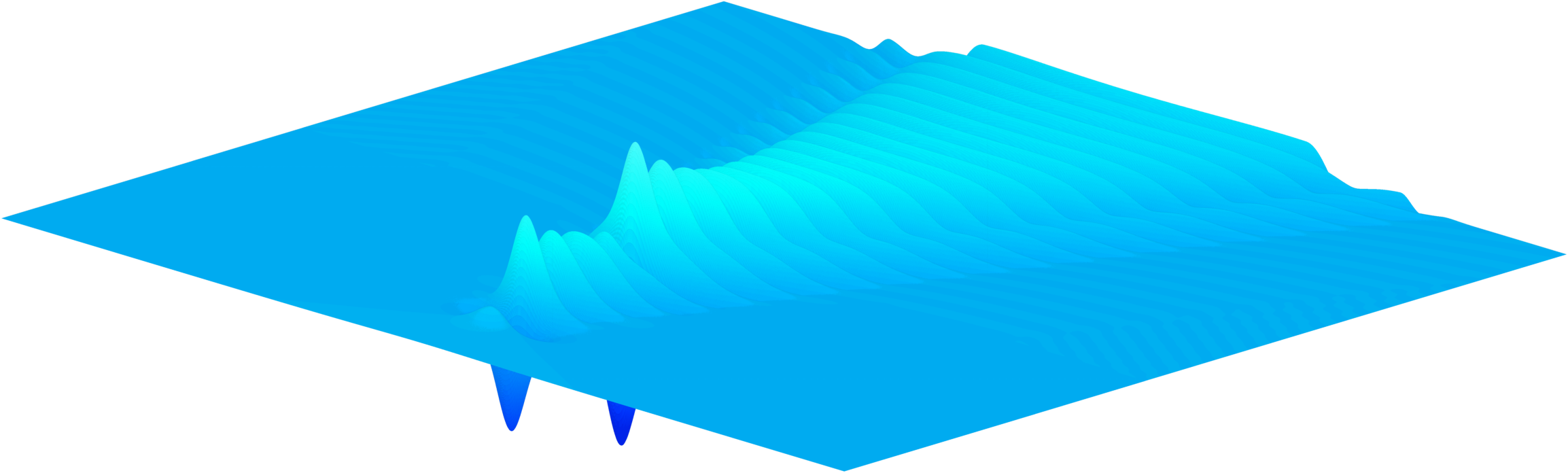}}
		\caption{Linear, separation distance = $9.12$}
		\label{}
	\end{subfigure}
	\begin{subfigure}[t]{0.3\textwidth}
		\centering
		\includegraphics[width=\textwidth]{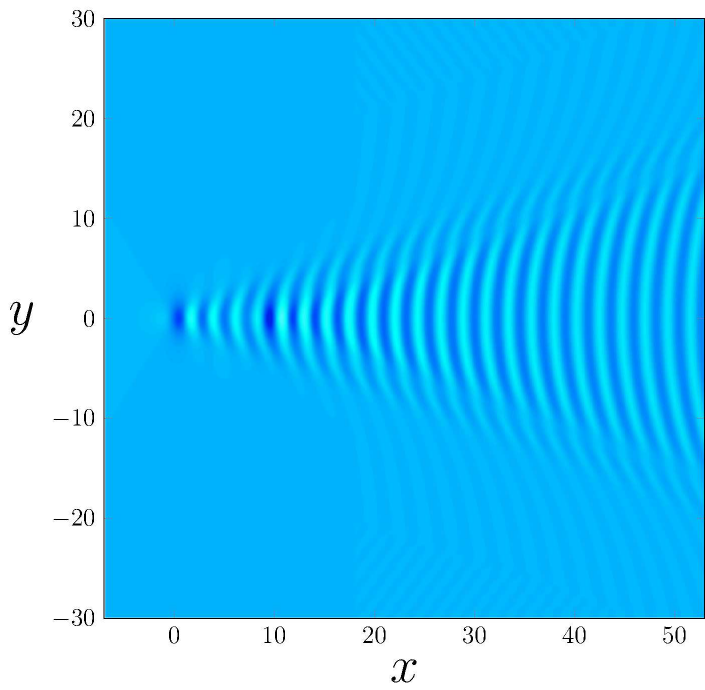}
		\caption{Plan view}
		\label{}
	\end{subfigure}
	\\
	\begin{subfigure}[t]{0.6\textwidth}
		\centering
		\raisebox{2em}{\includegraphics[width=\textwidth]{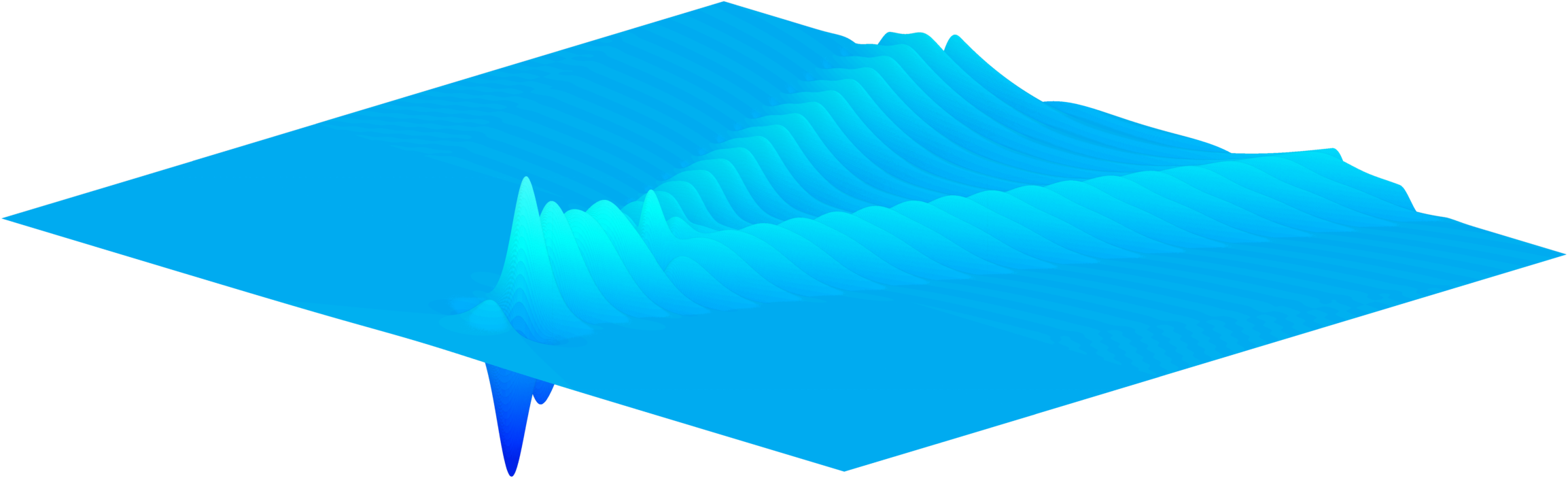}}
		\caption{Linear, separation distance = $10.26$}
		\label{}
	\end{subfigure}
	\begin{subfigure}[t]{0.3\textwidth}
		\centering
		\includegraphics[width=\textwidth]{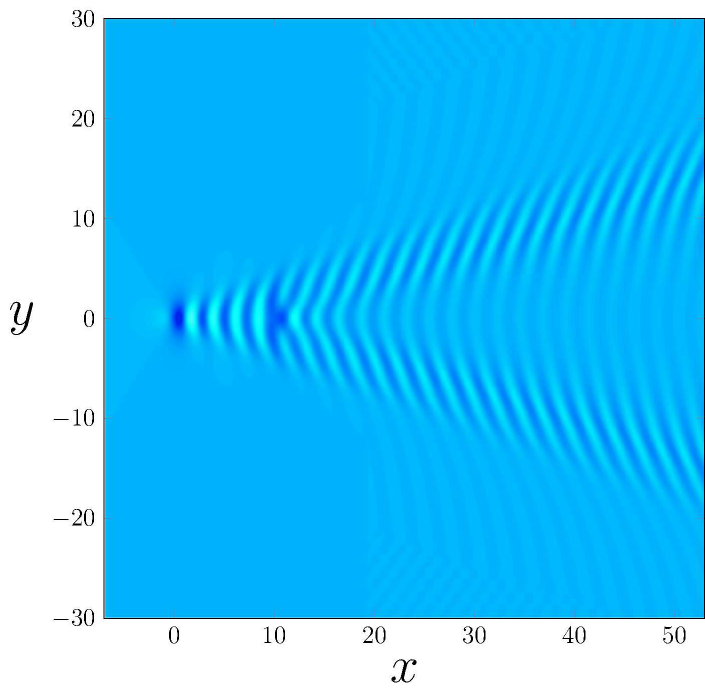}%
		\caption{Plan view}
		\label{}
	\end{subfigure}
	\\
	\begin{subfigure}[t]{0.6\textwidth}
		\centering
		\raisebox{2em}{\includegraphics[width=\textwidth]{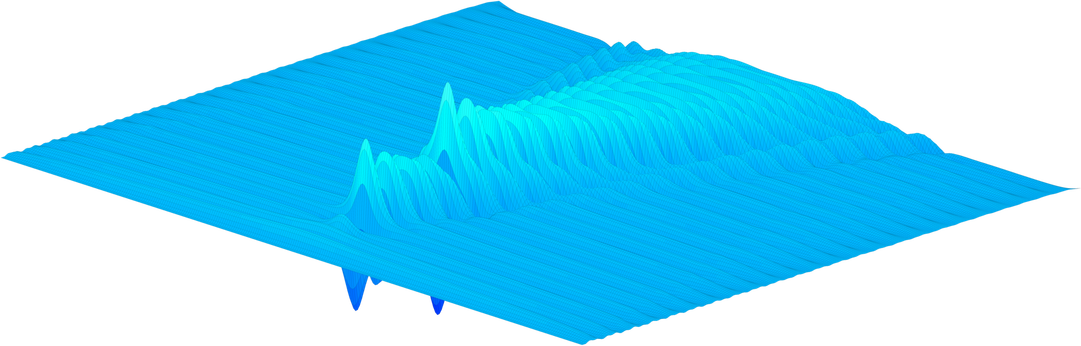}}
		\caption{Nonlinear, separation distance = $9.92$}
		\label{}
	\end{subfigure}
	\begin{subfigure}[t]{0.3\textwidth}
		\centering
		\includegraphics[width=\textwidth]{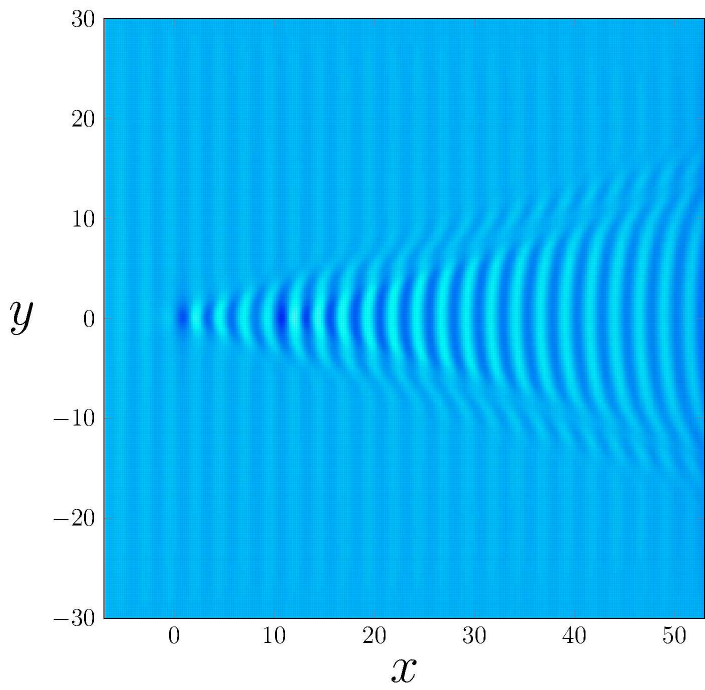}%
		\caption{Plan view}
		\label{}
	\end{subfigure}
	\\
	\begin{subfigure}[t]{0.6\textwidth}
		\centering
		\raisebox{2em}{\includegraphics[width=\textwidth]{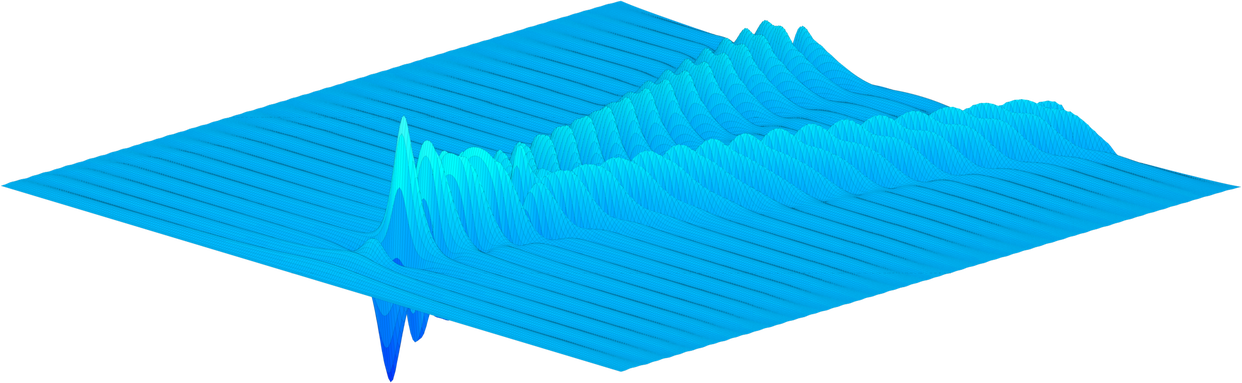}}
		\caption{Nonlinear, separation distance = $11.16$}
		\label{}
	\end{subfigure}
	\begin{subfigure}[t]{0.3\textwidth}
		\centering
		\includegraphics[width=\textwidth]{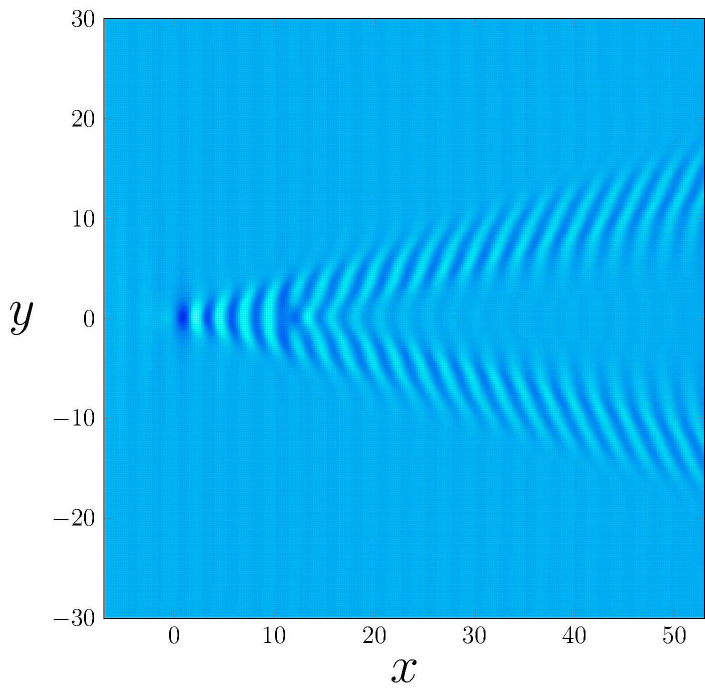}
		\caption{Plan view}
		\label{}
	\end{subfigure}
	\caption{{ Free-surface profiles for linear and nonlinear flow past two submerged bumps positioned inline along the $x$-axis, calculated for $\epsilon = 0.1$, $\delta = 0.5$ and $F = 0.6$.  The separation distance between the humps is chosen so that the wave profiles exhibit constructive interference (linear in (a)-(b) and nonlinear in (e)-(f)) and destructive interference (linear in (c)-(d) and nonlinear in (g)-(h)) of the transverse waves.  The nonlinear solutions were generated for $-7 \leq x \leq 53$ using a $281 \times 281$ mesh with $\Delta x = \Delta y = 0.2143$.  Note the centerline plots for these solutions is provided in Figure~\ref{fig:constructive_vs_destructive_centrelines}.}}
	\label{fig:inline_bumps}
\end{figure*}

{ We now consider the situation in which there are two bumps positioned inline along the $x$-axis.  Four free surface profiles for this configuration are presented in Figure~\ref{fig:inline_bumps}, again for $F=0.6$.  The profile in Figure~\ref{fig:inline_bumps}(a)-(b) is a linear solution computed for two bumps that are positioned 4 wavelengths apart, giving rise to constructive interference of the transverse waves.  In contrast, the profile in Figure~\ref{fig:inline_bumps}(c)-(d) is a linear solution computed for two bumps that are positioned 4.5 wavelengths apart, leading to destructive interference.  Here a single wavelength is given by $2\pi/k$ where $k$ is the real positive root of the linear dispersion relation $kF^2-\tanh k=0$ (for $F=0.6$ this linear wavelength is approximately $2.28$).  As the transverse waves decay in amplitude (like $x^{-1/2}$ for $x\gg 1$) the interference is not absolute (for example, the waves do not completely cancel each other out in the new Figure 8(c)-(d)).  However, the effects are strong and for the case of constructive interference, we see the resulting transverse waves downstream become larger than the divergent waves, which is unusual for this Froude number.}

\begin{figure*}
	\centering
	\begin{subfigure}[t]{0.45\textwidth}
		\centering
		\includegraphics[width=\linewidth]{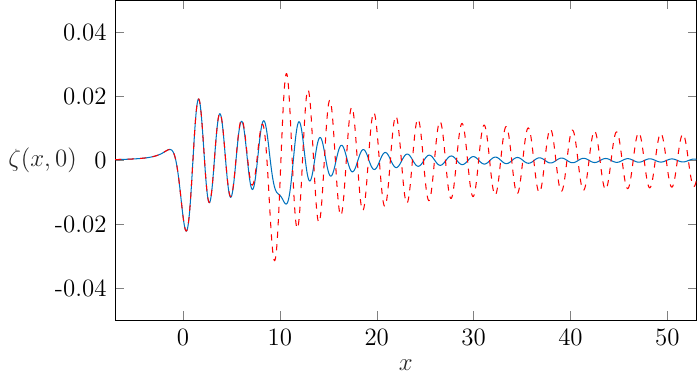}%
		\caption{Linear}
	\end{subfigure}
	\begin{subfigure}[t]{0.45\textwidth}
		\centering
		\includegraphics[width=\linewidth]{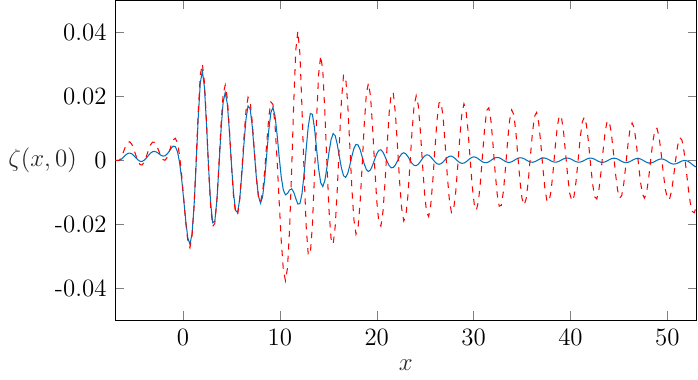}
		\caption{Nonlinear}
	\end{subfigure}
	\caption{{ Centreline plots exhibiting both destructive (blue solid) and constructive (red dashed) interference along the $y$-axis, computed for $\epsilon = 0.1$, $\delta = 0.5$ and $F = 0.6$, as in Figure~\ref{fig:inline_bumps}.  The linear solutions are in part (a) while the nonlinear solutions are in part (b).}}
	\label{fig:constructive_vs_destructive_centrelines}
\end{figure*}

{ Analogous nonlinear profiles are presented in Figure~\ref{fig:inline_bumps}(e)-(f) and Figure~\ref{fig:inline_bumps}(g)-(h).  In this nonlinear case there is no obvious wavelength as the linear theory no longer applies.  Instead, we have computed these solutions for bumps that are roughly 4 and 4.5 wavelengths apart, where a single wavelength is estimated from a nonlinear solution with a single hump.  Again, the visual differences between the two cases are clear from these figures.  To support these results, we have presented in Figure~\ref{fig:constructive_vs_destructive_centrelines} centreline plots which more clearly show the effects of (transverse) wave interference.  Similar plots can be generated for bumps that are separated by any number of wavelengths and the wave profiles exhibit similar properties.  These results are reminiscent of the wave interference effects that occur at the stern of a steadily moving ship, although in that context it is also of interest to study interference by divergent waves (\cite{noblesse14,zhang15a,zhu15}).}

\subsection{Flow past a crater}
\begin{figure*}
\centering
\begin{subfigure}[t]{0.6\linewidth}
\raisebox{2.5em}{\includegraphics[width=\linewidth]{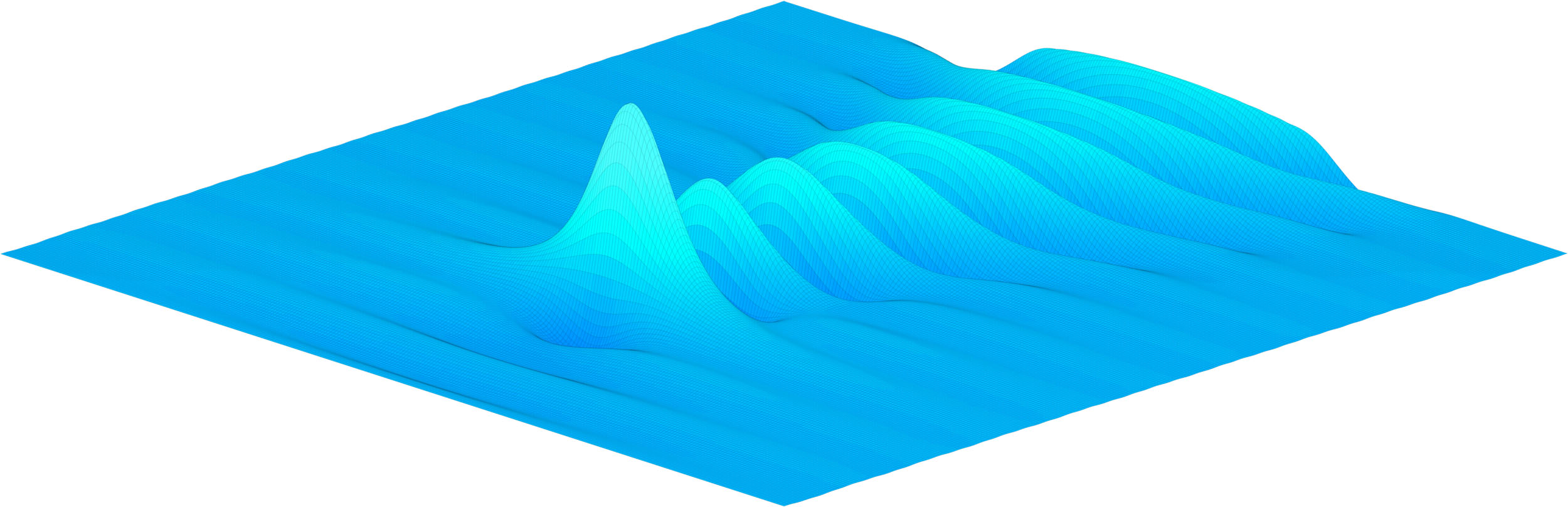}}
\caption{Single crater, $\epsilon=-0.1$}
\label{fig:crater1}
\end{subfigure}
\begin{subfigure}[t]{0.3\linewidth}
\includegraphics[width=\linewidth]{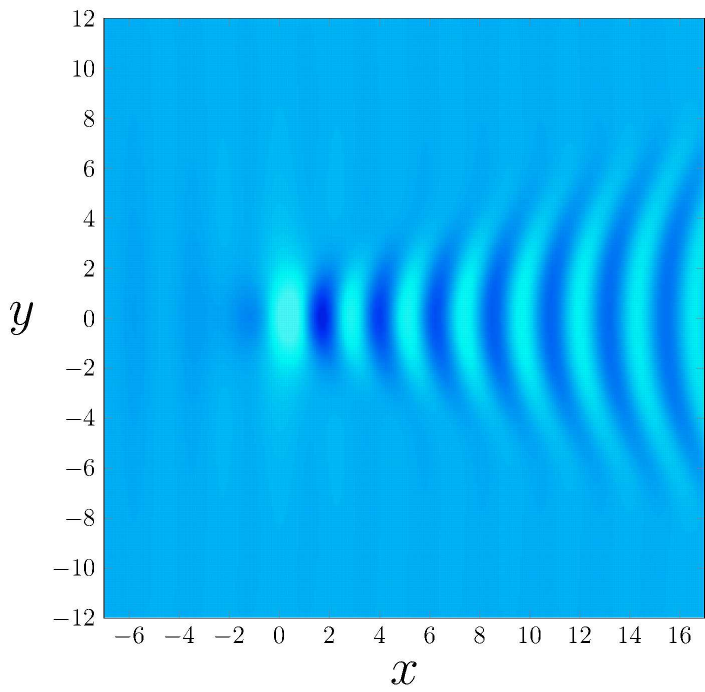}
\caption{Plan view}
\label{fig:crater2}
\end{subfigure}
\\
\begin{subfigure}[t]{0.6\linewidth}
\centering
\raisebox{2.5em}{\includegraphics[width=\linewidth]{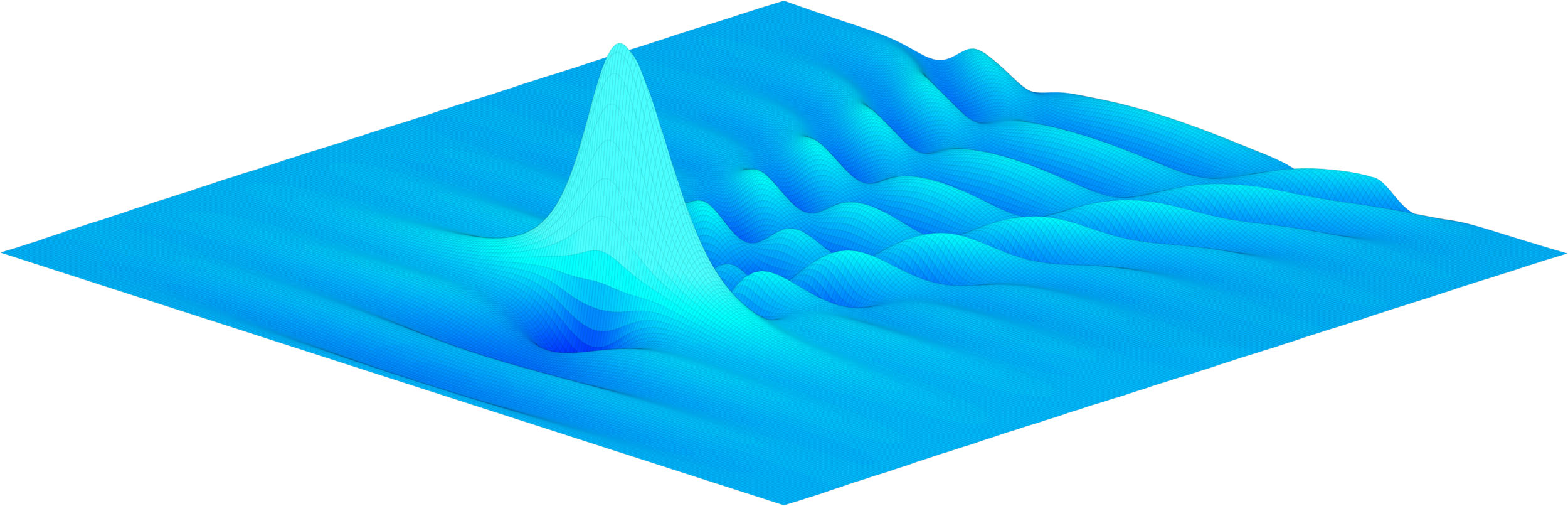}}
\caption{Single crater, $\epsilon=-0.62$}
\label{fig:crater3}
\end{subfigure}
\begin{subfigure}[t]{0.3\linewidth}
\centering
\includegraphics[width=\linewidth]{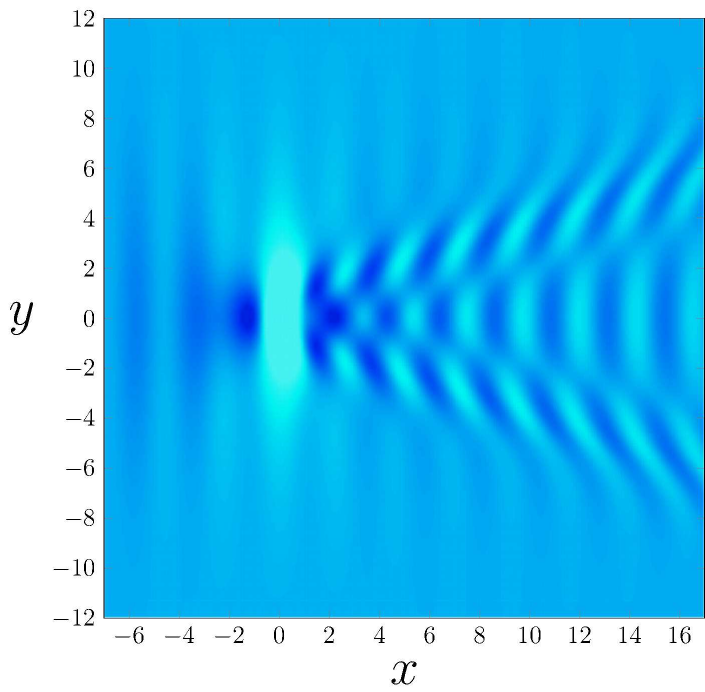}
\caption{Plan view}
\label{fig:crater4}
\end{subfigure}
\caption{Nonlinear solutions for subcritical flow past a submerged crater centred at the origin. Solutions were computed using a $281\times 281$ mesh with $\Delta x = \Delta y = 0.0857$, $x_{1}=-7$, $\delta=0.5$, $F=0.6$, $\epsilon=-0.62$. The solutions shows that for the near linear case, the wave patterns are the inverse of the solution for flow past a bump, however, for the more nonlinear case the first wave crest is much higher than the rest of the waves and the transverse waves appear to be out of phase with the divergent waves.  As an indication of the relative vertical scales in these plots, the maximum height for the solution in (a)-(b) is 0.020 while the maximum height for the solution in (c)-(d) is 0.073.}
\label{fig:crater}
\end{figure*}

We now turn our attention to flow past an isolated crater, which is easy to implement with our numerical scheme (see \citet{Wade2017,Zhang1996b} for the two-dimensional analogue). In particular, we consider flow over an inverted Gaussian, which is (\ref{eq:bottom}) with $\epsilon<0$.   Figure~\ref{fig:crater}(a)-(b) shows surface profiles for $\epsilon=-0.1$, which is a relatively shallow crater, while Figure~\ref{fig:crater}(c)-(d) is for $\epsilon=-0.62$, a relatively deep crater.  The solution for $\epsilon=-0.1$ has a wave pattern that is very similar to the linear solution (not shown) and is also an almost-exact reflection of the surface for $\epsilon=0.1$ (see Figure~\ref{fig:subcritical}(c)-(d)).  On the other hand, the highly nonlinear solution for $\epsilon=-0.62$ is interesting because its wave pattern appears very different from both the linear solution and the nonlinear solution for flow over a standard Gaussian.  For example, even though this is a highly nonlinear solution, the divergent waves do not resemble the steep dorsal fins we see in Figure~\ref{fig:subcritical}(e).

\section{Discussion}\label{sec:discussion}

We have conducted a numerical investigation into the problem of three-dimensional steady free-surface flow over an arbitrary bottom topography.  By applying the boundary integral framework outlined by \citet{Forbes1}, we have been able to compute fully nonlinear numerical solutions for a range of localised bottom obstructions.  Following our previous work \citep{Pethiyagoda14,pethiyagoda2014}, the approach we have followed is to develop a Jacobian-free Newton Krylov method which allows for many more grid points on the free surface than is normally used \citep{Parau1,Parau2011,Parau2005,Parau2,Parau2007a,Parau2007b}.  As such, we have been able to compute highly nonlinear solutions which show wave features that are not observed for the linear regime.

We now provide some further comments on the close relationship between the steady wave pattern that is generated by flow over an isolated bottom obstacle and the ship wake that forms behind the stern of a steadily moving ship.  This connection is particularly noteworthy because steady ship wakes have been the subject of renewed theoretical interest in recent years in the physics and applied mathematics community, for example in the context of determining the apparent wake angle \citep{darmon14,dias14,he14,miao15,noblesse14,pethiyagoda15,rabaud13}, understanding interference between bow and stern flows \citep{zhang15a,zhu15}, considering the effects of shear~\citep{ellingsen14,li16,smeltzer17} and unpicking the wave pattern via time-frequency analysis \citep{pethiyagoda17,pethiyagoda17b,torsvik15a}.

What is common to all of these papers is that the wave disturbance due to a ship is approximated by considering a steady moving pressure distribution applied to the free surface.  In an attempt to make the connection between these two problems clearer,  we noted in \S\ref{sec:linear} that, at least for the linear regime, flow past a given isolated pressure distribution gives exactly the same wake as flow past certain isolated bottom obstruction.  Thus we conclude there is a direct quantitative relationship between the two problems and many of the properties of ship wakes (such as apparent wake angle, wave interference, and time-frequency heat maps) carry over to the problem of flow past a bottom obstruction in the linear regime.

\begin{figure*}
\centering
\begin{subfigure}[t]{0.45\linewidth}
\centering
\raisebox{0em}{\includegraphics[width=\linewidth]{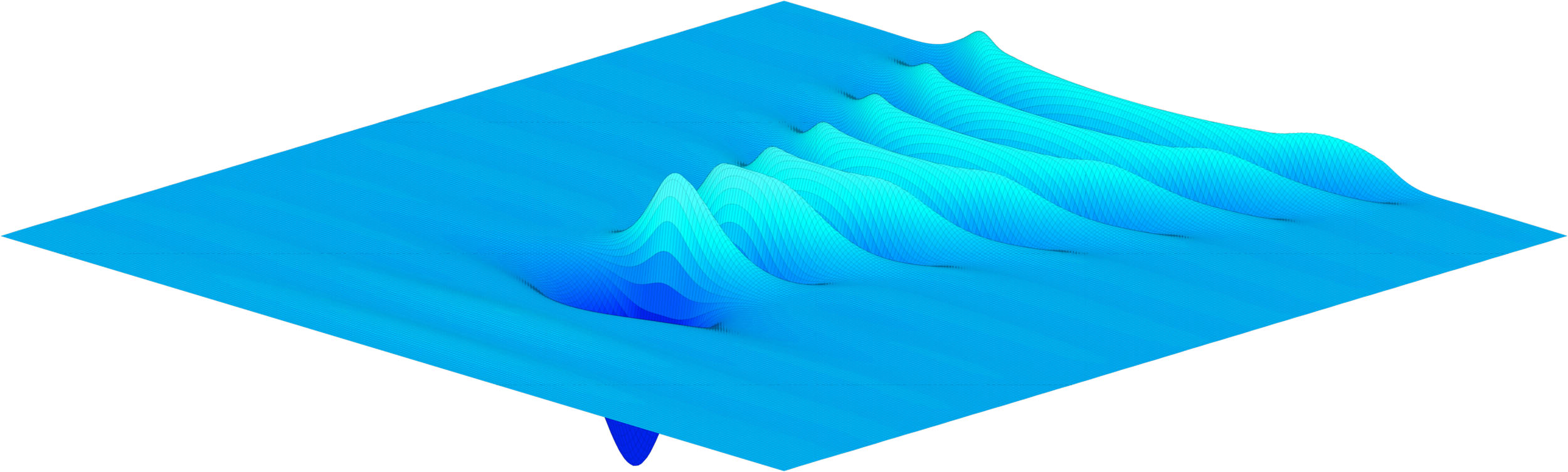}}
\end{subfigure}
\begin{subfigure}[t]{0.45\linewidth}
\includegraphics[width=\linewidth]{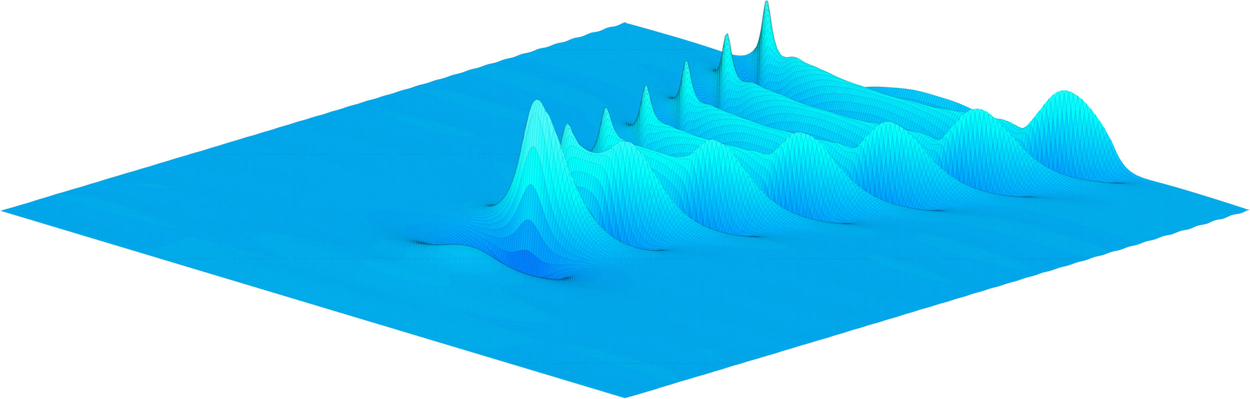}
\end{subfigure}
\caption{Nonlinear solutions for flow past a localised pressure distribution (\ref{eq:pressure}) [left] and an isolated bottom obstruction (\ref{eq:bottom}) [right] for $F=0.6$ on $-7\leq x\leq 17$, $-12\leq y\leq 12$ using a $281\times 281$ mesh with $\Delta x = \Delta y = 0.0857$, $\delta=0.5$, $\epsilon=0.285$.  For $\epsilon\ll 1$, these two problems have the same solution for the wake downstream; however, in this figure it is clear that in this highly nonlinear regime (large $\epsilon$), the computed surface wave patterns are very different.}
\label{fig:comparison}
\end{figure*}

The key further point to make here is that the nonlinear problems of flow past a pressure distribution and flow over a bottom obstruction are not the same.  As the size of the disturbance in each case increases, nonlinear effects become more pronounced.  We demonstrate this point in Figure~\ref{fig:comparison} by showing wave patterns for flow past a bottom obstruction (\ref{eq:bottom}) with $F=0.6$, $\epsilon=0.285$ and $\delta=0.5$ and flow past the pressure distribution
\begin{equation*}
p(x,y)=\frac{\epsilon \delta^{2}}{F^2} \int_{0}^{\infty} {k\, \mathrm{sech}(k) \mathrm{e}^{-\delta^{2} k^{2}/2} J_{0}\left(k\sqrt{x^{2}+y^{2}}\right)} \mathrm{d}k
\label{eq:pressure2}
\end{equation*}
with the same parameter values.  For this choice, the form of the wake downstream from the disturbance is the same in the linear regime $\epsilon \ll 1$.  However, as is clearly demonstrated in Figure~\ref{fig:comparison}, for a moderately large value of $\epsilon$ (which is a measure of nonlinearity), the surfaces are distinct. As the nonlinearity increases, the solution for flow past a bottom topography produces sharp crests on the divergent waves and the height of the transverse waves is much smaller than the divergent waves, contrasted with the flow past a pressure solution still looking linear.  Thus, in summary, the linearised problem of flow past one or a combination of isolated bumps is same as flow due to a number of isolated pressure patches on the surface (provided the precise form of the pressure is chosen according to (\ref{eq:pressure})), the latter being routinely used as a simple model for ship wakes.  However, the nonlinear versions of these problems are very different and only through a fully nonlinear numerical scheme can these differences be highlighted.

{  A benefit of our numerical scheme is the ease with which it can be extended to include other physical effects.  For example, if the effects of surface tension are included on the free surface, the only change to the governing equations (\ref{eq:laplace})-(\ref{eq:new_finite_farfield3}) is to add a higher-order term in Bernoulli's equation to represent the surface tension parameter multiplied by mean curvature (\cite{Parau2007b}).  For our purposes, we need to consider the effects of this extra term on the Jacobian of the system.  It turns out that the top left block (of the nine blocks) of the Jacobian is altered by surface tension, with extra non-zero entries appearing near the main diagonal.  As a consequence, we expect that our approach of using the Jacobian from the linear version of the problem as a preconditioner will carry over, and thus we should be able adapt our scheme to compute accurate solutions with gravity-capillary waves.  We leave this topic for future work.}

We close by mentioning that our work can be adapted in many ways to further explore free-surface flows over bottom topographies.  As mentioned in the Introduction, there is a plethora of published research for the two-dimensional case and, with our scheme, many extensions to three dimensions are now possible.  Furthermore, the shape and properties of highly nonlinear steady ship waves are still not well understood, for the most part due to an absence of accurate numerical solutions.  Our numerical approach provides an opportunity for computational results in this area of interest.

\section{Acknowledgements}\label{sec:acknowledgements}
S.W.M. acknowledges the support of the Australian Research Council via the Discovery Project DP140100933. The authors acknowledge the computational resources and technical support provided by the High Performance Computing and Research Support group, Queensland University of Technology.  We are grateful to the three anonymous referees for their insightful comments.

\appendix
\section{Banding the preconditioner}\label{app:banded}
Storing the full $\mathbf{P}$ requires a large amount of memory. To overcome this problem $\mathbf{P}$ can be split into blocks, which can each use different storage methods. The sparse blocks, $\mathbf{P}_{1,1}$, $\mathbf{P}_{1,2}$, $\mathbf{P}_{1,3}$, $\mathbf{P}_{2,2}$ and $\mathbf{P}_{3,3}$, can utilise sparse storage methods while the dense blocks can be stored separately. Alternatively, the dense blocks can not be stored at all and can just be formed whenever the preconditioner is being applied. This leads to a smaller memory requirement but leads to a large increase in runtime.

\begin{figure*}
\centering
\begin{subfigure}[t]{0.45\linewidth}
\centering
\includegraphics[width=0.9\linewidth]{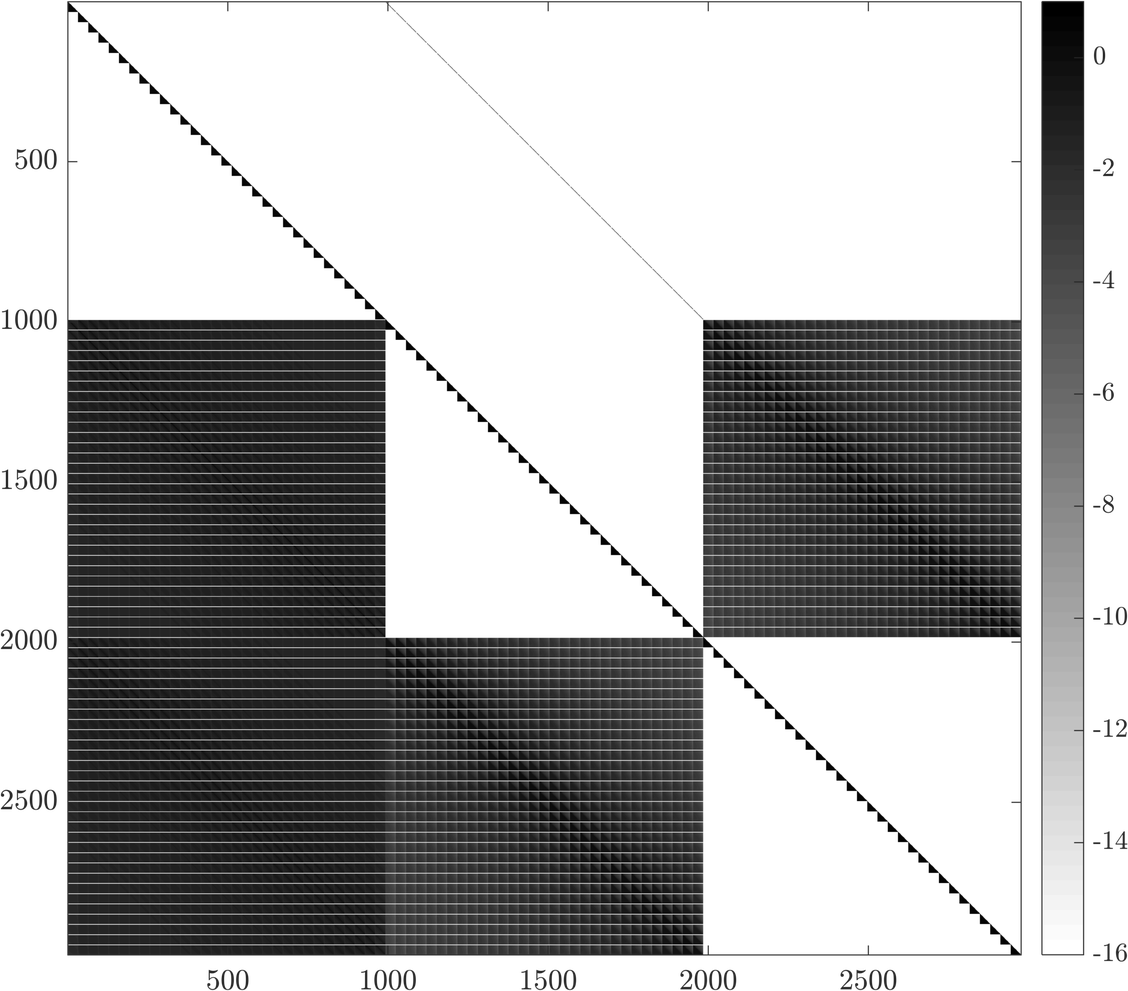}
\caption{Full linear Jacobian, $\mathbf{\tilde{J}}$}
\label{fig:jac_full}
\end{subfigure}
\begin{subfigure}[t]{0.45\linewidth}
\centering
\includegraphics[width=0.9\linewidth]{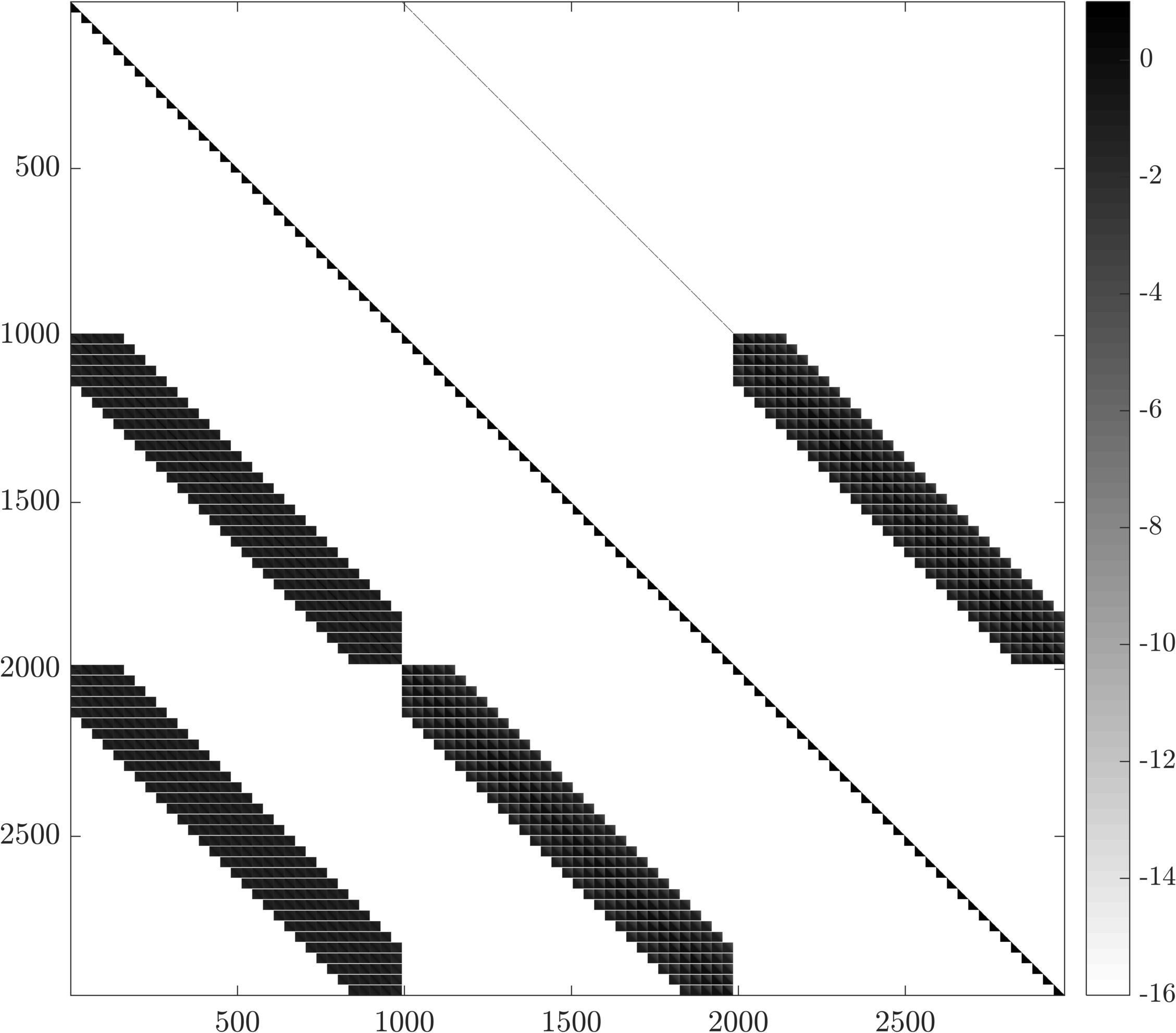}
\caption{Banded preconditioner, $\mathbf{P}$}
\label{fig:jac_band}
\end{subfigure}
\\
\begin{subfigure}[t]{0.45\linewidth}
\centering
\includegraphics[width=0.9\linewidth]{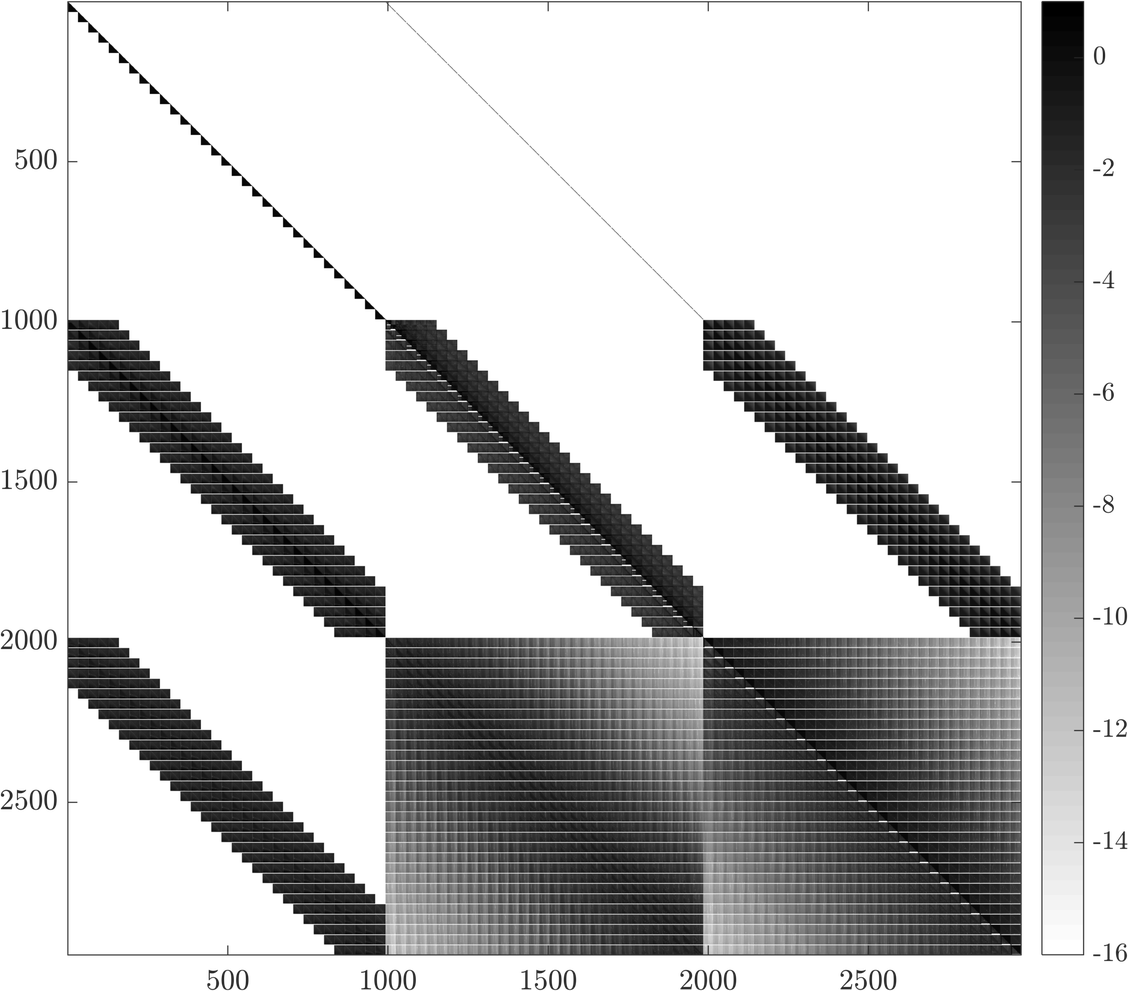}
\caption{Factored preconditioner}
\label{fig:jac_band_lu}
\end{subfigure}
\caption{Visualisations of the full, block banded, block LU block banded, and block iLU block banded preconditioner on a $31 \times 31$ mesh with $\Delta x = \Delta y = 0.4$, $F=0.6$ and a single Gaussian bump on the bottom surface with scale parameter $\epsilon=0.1$. (a) shows the full Jacobian from the linear problem. (b) shows the banded approximation of (a), showing a great reduction nonzero values resulting in less memory required for storage. (c) shows the block LU factorisation of (b), note that only the centre and right blocks in the third row become dense after the block LU factorisation. By banding the blocks before factorising, the number of dense blocks has been reduced from six down to two.}
\label{fig:jacobian_banding}
\end{figure*}

For use in the JFNK algorithm, it is beneficial for the preconditioner to be factorised such that is can be applied faster when necessary. Since $\mathbf{P}$ is already split into blocks, a block LU factorisation can be utilised. This factorisation treats $\mathbf{P}$ as a $3\times3$ matrix and applies the factorisation: $\mathbf{P} = \mathbf{L}\mathbf{U}$, where $\mathbf{L}$ and $\mathbf{U}$ are $3\times3$ block lower and upper triangular matrices respectively with $(N+1)M\times(N+1)M$ identity matrices along the main diagonal of $\mathbf{L}$.
The nine factored blocks are stored separately. While this factorisation makes applying the preconditioner more effective, the process leads to more dense blocks in storage. To reduce the memory required to store all the blocks the dense blocks can be approximated. Figure \ref{fig:jacobian_compare}(b) shows that in the dense blocks the magnitudes of the values decay away from the main diagonal, leading to the idea that a banded approximation could be used. Figure \ref{fig:jacobian_banding} shows a visualisation of the full and banded $\mathbf{P}$. The banded storage utilised takes block bands, that is, it takes $b$ $(N+1)\times (N+1)$ blocks in each row. This is done instead of regular banding because of the block structure in the matrix. Keeping these blocks ensures that all of the values that correspond to an entire slice of the $y$-axis are kept in $\mathbf{P}$. From the visualisation it is clear that banding the dense blocks will reduce the memory requirement for storing $\mathbf{P}$ as there are far fewer elements being stored. Table \ref{table:memory} shows the details of the memory savings achieved by banding the dense blocks of $\mathbf{P}$.

\begin{table}
\centering
\begin{tabular}{ccc} \hline
Mesh size & Full storage & Banded, $b=\frac{M}{10}$ \\ \hline
$31\times 31$ & $30.8$ MB & $16.2$ MB \\
$61\times 61$ & $442.2$ MB & $206.9$ MB \\
$151\times 151$ & $59.0$ GB & $9.6$ GB \\ \hline
\end{tabular}
\caption{Table showing the approximate memory used to store the preconditioner for particular mesh sizes. As the mesh size increases, the amount of memory required for the full preconditioner increases rapidly. The memory savings from using the banded preconditioner are more pronounced for finer meshes as the bandwidth is taken to be a fraction of the number of mesh points in the $y$ direction.}
\label{table:memory}
\end{table}

\newpage
\section{Algebraic equations for linearised problem}\label{sec:linear_jacobian}
The linear problem presented in \S\ref{sec:linear} is discretised in the same manner as the nonlinear problem in \S\ref{sec:numerical_collocation} and evaluated on the half mesh points $(x^*_k,y^*_l)=((x_k+x_{k+1})/2,y_l)$. The resulting vector function, $\mathbf{F}$, is:
\begin{align}
\mathbf{F}_{1(k,l)} &= \phi^{*}_{x(k,l)} + \frac{\zeta^{*}_{(k,l)}}{F_{H}^{2}} - 1, \\
\mathbf{F}_{2(k,l)} &= \sum_{i=1}^{N} \sum_{j=1}^{M} w(i,j) \left\{ \left(\zeta_{x(i,j)} - \zeta_{x(k,l)}^{*}\right) K_{5(i,j,k,l)} \right. \nonumber \\
& \quad {} + \left. \beta_{x(i,j)} K_{7(i,j,k,l)} + (\psi_{(i,j)} - x_i) K_{6(i,j,k,l)} \right\}\nonumber \\
& \quad {} + \zeta_{x(k,l)}^{*} \iint K_{5} \text{ d}x\text{ d}y - 2\pi \left(\phi_{(k,l)}^{*} - x_{k}^{*}\right), \\
\mathbf{F}_{3(l)}   &= x_{1} \phi_{x(1,l)} + n \phi_{(1,l)} - x_{1}(n+1), \\
\mathbf{F}_{4(l)}   &= \frac{x_{1}}{\Delta x} \phi_{x(2,l)} + \left( n - \frac{x_{1}}{\Delta x}\right) \phi_{x(1,l)} - n, \\
\mathbf{F}_{5(l)}   &= x_{1} \zeta_{x(1,l)} + n \zeta_{(1,l)}, \\
\mathbf{F}_{6(l)}   &= \frac{x_{1}}{\Delta x} \zeta_{x(2,l)} + \left( n - \frac{x_{1}}{\Delta x} \right) \zeta_{x(1,l)}, \\
\mathbf{F}_{7(k,l)} &= \sum_{i=1}^{N} \sum_{j=1}^{M} w(i,j) \Big\{ \zeta_{x(i,j)} K_{7(i,j,k,l)} \nonumber \\
& \quad {} + \left( -\beta_{x(k,l)}^{*} + \beta_{x(i,j)}\right) K_{5(i,j,k,l)} \nonumber \\
& \quad {} + (\phi_{(i,j)} - x_i) K_{6(i,j,k,l)} \Big\} \nonumber \\
& \quad {} + \beta_{x(k,l)}^{*} \iint K_{5} \text{ d}x\text{ d}y \nonumber \\
& \quad {} - 2\pi \left(\psi_{(k,l)}^{*} - x_{k}^{*}\right), \\
\mathbf{F}_{8(l)}   &= x_{1} \psi_{x(1,l)} + n \psi_{(1,l)} - x_{1}(n+1), \\
\mathbf{F}_{9(l)}   &= \frac{x_{1}}{\Delta x} \psi_{x(2,l)} + \left( n - \frac{x_{1}}{\Delta x}\right) \psi_{x(1,l)} - n, \\
\end{align}
for $k=1,2,...,N-1$, $l=1,2,...,M$ and
\begin{align}
K_{5(i,j,k,l)} = \left((x_{i} - x^*_{k})^{2} + (y_{j} - y^*_{l})^{2}\right)^{-1/2}, \\
K_{6(i,j,k,l)} = \left((x_{i} - x^*_{k})^{2} + (y_{j} - y^*_{l})^{2} + 1\right)^{-3/2}, \\
K_{7(i,j,k,l)} = \left((x_{i} - x^*_{k})^{2} + (y_{j} - y^*_{l})^{2} + 1\right)^{-1/2}.
\end{align}
The order of the functions in the vector function is:
\begin{align}
\mathbf{F} = \begin{bmatrix}
\mathbf{F}_{5(1)} \\
\mathbf{F}_{6(1)} \\
\mathbf{F}_{1(1,1)} \\
\mathbf{F}_{1(2,1)} \\
\vdots \\
\mathbf{F}_{1(N-1,1)} \\
\mathbf{F}_{5(2)} \\
\vdots \\
\mathbf{F}_{1(N-1,M)} \\
\mathbf{F}_{3(1)} \\
\mathbf{F}_{4(1)} \\
\mathbf{F}_{2(1,1)}
\vdots \\
\mathbf{F}_{2(N-1,M)} \\
\mathbf{F}_{8(1)} \\
\mathbf{F}_{9(1)} \\
\mathbf{F}_{7(1,1)} \\
\vdots \\
\mathbf{F}_{7(N-1,M)}
\end{bmatrix}.
\end{align}
The simplicity of the linear discretised problem allows for the Jacobian to be computed analytically. This is done by differentiating the equations with respect to each of the unknowns.
\begin{align}
\frac{\mathrm{d}\mathbf{F}}{\mathrm{d}\mathbf{u}} &= \begin{bmatrix}
\pardiff{\mathbf{F}_{(1)}}{\mathbf{u}_{1}} & \pardiff{\mathbf{F}_{(1)}}{\mathbf{u}_{2}} & \ldots & \pardiff{\mathbf{F}_{(1)}}{\mathbf{u}_{3(N+1)M}} \\
\pardiff{\mathbf{F}_{(2)}}{\mathbf{u}_{1}} & \pardiff{\mathbf{F}_{(2)}}{\mathbf{u}_{2}} & \ldots & \pardiff{\mathbf{F}_{(2)}}{\mathbf{u}_{3(N+1)M}} \\
\vdots & \vdots & \ddots & \vdots \\
\pardiff{\mathbf{F}_{(3(N+1)M)}}{\mathbf{u}_{1}} & \pardiff{\mathbf{F}_{(3(N+1)M)}}{\mathbf{u}_{2}} & \ldots & \pardiff{\mathbf{F}_{(3(N+1)M)}}{\mathbf{u}_{3(N+1)M}}
\end{bmatrix}
\end{align}
The derivatives of the function are:
\begin{align}
\pardiff{\mathbf{F}_{1(k,l)}}{\phi_{(1,m)}} &= 0, \\
\pardiff{\mathbf{F}_{1(k,l)}}{\phi_{x(n,m)}} &= \begin{cases}
\frac{1}{2} \quad \text{for } n=k, \; n=k+1, \; m=l, \\
0 \quad \text{otherwise}.
\end{cases}, \\
\pardiff{\mathbf{F}_{1(k,l)}}{\zeta_{(1,m)}} &= \begin{cases}
\frac{1}{F_{H}^{2}} \quad \text{for } m=l, \\
0 \quad \text{otherwise}.
\end{cases}, \\
\pardiff{\mathbf{F}_{1(k,l)}}{\zeta_{x(n,m)}} &= \begin{cases}
- \frac{\Delta x}{2 F_{H}^{2}} \quad \text{for } n=1, \; m=l, \\
\frac{\Delta x}{F_{H}^{2}} \quad \text{for } 1<n<k, \; m=l, \\
\frac{3 \Delta x}{4 F_{H}^{2}} \quad \text{for } n=k, \; m=l, \\
\frac{\Delta x}{4 F_{H}^{2}} \quad \text{for } n=k+1, \; m=l, \\
0, \quad \text{otherwise}.
\end{cases}, \\
\pardiff{\mathbf{F}_{1(k,l)}}{\psi_{x(n,m)}} &= 0, \\
\pardiff{\mathbf{F}_{2(k,l)}}{\phi_{(1,m)}} &= \begin{cases}
- 2\pi \quad \text{for } m=l, \\
0 \quad \text{otherwise}.
\end{cases}, \\
\pardiff{\mathbf{F}_{2(k,l)}}{\phi_{x(n,m)}} &= \begin{cases}
- \frac{\pi}{2} \Delta x \quad \text{for } n=1, \; k=1, \; m=l, \\
- \pi \Delta x \quad \text{for } n=1, \; k>1, \; m=l, \\
- 2 \pi \Delta x \quad \text{for } 1<n<k, \; m=l, \\
- \frac{3 \pi \Delta x}{2} \quad \text{for } n=k, \; m=l, \\
- \frac{\pi \Delta x}{2} \quad \text{for } n=k+1, \; m=l, \\
0 \quad \text{otherwise}.
\end{cases}, \\
\pardiff{\mathbf{F}_{2(k,l)}}{\zeta_{(1,m)}} &= 0, \\
\pardiff{\mathbf{F}_{2(k,l)}}{\zeta_{x(n,m)}} &= \begin{cases}
\frac{1}{2} \sum\limits_{i=1}^{N} \sum\limits_{j=1}^{M} w(i,j) K_{5(i,j,k,l)} \\
\quad {} - \frac{1}{2} \iint K_{5} \\
\quad {} - w(n,m)K_{5(n,m,k,l)} \\
\qquad \qquad \qquad \qquad \text{for } n=k, \; m=l, \\
-w(n,m) K_{5(n,m,k,l)} \quad \text{otherwise}.
\end{cases}, \\
\pardiff{\mathbf{F}_{2(k,l)}}{\psi_{(1,m)}} &= 0, \\
\pardiff{\mathbf{F}_{2(k,l)}}{\psi_{x(n,m)}} &= \begin{cases}
\sum\limits_{i=2}^{N} \left\{ \frac{\Delta x}{2} w(i,m) K_{6(i,m,k,l)} \right\} \quad \text{for } n=1, \\
\frac{\Delta x}{2} w(n,m) K_{6(n,m,k,l)} \\
\quad {} + \Delta x \sum\limits_{i=n+1}^{N} \left\{ w(i,m) K_{6(i,m,k,l)} \right\} \\
\qquad \qquad \qquad \qquad \text{for } 1<n<N, \\
\frac{\Delta x}{2} w(n,m) K_{6(n,m,k,l)} \quad \text{for } n=N.
\end{cases}, \\
\pardiff{\mathbf{F}_{3(l)}}{\phi_{(1,m)}} &= n \quad \text{for } m=l, \\
\pardiff{\mathbf{F}_{3(l)}}{\phi_{x(1,m)}} &= \begin{cases}
x_{1} \quad \text{for } m=l, \\
0 \quad \text{otherwise}.
\end{cases}, \\
\pardiff{\mathbf{F}_{3(l)}}{\zeta_{(1,m)}} &= 0, \\
\pardiff{\mathbf{F}_{3(l)}}{\zeta_{x(n,m)}} &= 0, \\
\pardiff{\mathbf{F}_{3(l)}}{\psi_{x(k,l)}} &= 0, \\
\pardiff{\mathbf{F}_{4(l)}}{\phi_{(1,l)}} &= 0, \\
\pardiff{\mathbf{F}_{4(l)}}{\phi_{x(n,m)}} &= \begin{cases}
n - \frac{x_{1}}{\Delta x} \quad \text{for } n=1, \; m=l, \\
\frac{x_{1}}{\Delta x} \quad \text{for } n=2, \; m=l, \\
0 \quad \text{otherwise}.
\end{cases}, \\
\pardiff{\mathbf{F}_{4(l)}}{\zeta_{(1,m)}} &= 0, \\
\pardiff{\mathbf{F}_{4(l)}}{\zeta_{x(n,m)}} &= 0, \\
\pardiff{\mathbf{F}_{4(l)}}{\psi_{x(n,m)}} &= 0, \\
\pardiff{\mathbf{F}_{5(l)}}{\phi_{(1,m)}} &= 0, \\
\pardiff{\mathbf{F}_{5(l)}}{\phi_{x(n,m)}} &= 0, \\
\pardiff{\mathbf{F}_{5(l)}}{\zeta_{(1,m)}} &= \begin{cases}
n \quad \text{for } m=l, \\
0 \quad \text{otherwise}.
\end{cases}, \\
\pardiff{\mathbf{F}_{5(l)}}{\zeta_{x(n,m)}} &= \begin{cases}
x_{1} \quad \text{for } n=1, \; m=l, \\
0 \quad \text{otherwise}.
\end{cases}, \\
\pardiff{\mathbf{F}_{5(l)}}{\psi_{x(k,l)}} &= 0, \\
\pardiff{\mathbf{F}_{6(l)}}{\phi_{(1,m)}} &= 0, \\
\pardiff{\mathbf{F}_{6(l)}}{\phi_{x(n,m)}} &= 0, \\
\pardiff{\mathbf{F}_{6(l)}}{\zeta_{(1,m)}} &= 0, \\
\pardiff{\mathbf{F}_{6(l)}}{\zeta_{x(n,m)}} &= \begin{cases}
n - \frac{x_{1}}{\Delta x} \quad \text{for } n=1, \; m=l, \\
\frac{x_{1}}{\Delta x} \quad \text{for } n=2, \; m=l, \\
0 \quad \text{otherwise}.
\end{cases}, \\
\pardiff{\mathbf{F}_{6(l)}}{\psi_{x(n,m)}} &= 0, \\
\pardiff{\mathbf{F}_{7(k,l)}}{\phi_{(1,m)}} &= 0, \\
\pardiff{\mathbf{F}_{7(k,l)}}{\phi_{x(n,m)}} &= \begin{cases}
\sum\limits_{i=2}^{N} \left\{ \frac{\Delta x}{2} w(i,m) K_{6(i,m,k,l} \right\} \\
\qquad \qquad \qquad \qquad \text{for } n=1, \\
\frac{\Delta x}{2} w(n,m) K_{6(n,m,k,l)} \\
\quad {} + \Delta x \sum\limits_{i=n+1}^{N} \left\{ w(i,m) K_{6(i,m,k,l)} \right\} \\
\qquad \qquad \qquad \qquad \text{for } 1<n<N, \\
\frac{\Delta x}{2} w(n,m) K_{6(n,m,k,l)} \quad \text{for } n=N.
\end{cases}, \\
\pardiff{\mathbf{F}_{7(k,l)}}{\zeta_{(1,m)}} &= 0, \\
\pardiff{\mathbf{F}_{7(k,l)}}{\zeta_{x(n,m)}} &= \begin{cases}
-w(n,m) K_{5} \quad \text{for } m=l, \\
0 \quad \text{otherwise}.
\end{cases}, \\
\pardiff{\mathbf{F}_{7(k,l)}}{\psi_{x(n,m)}} &= \begin{cases}
- \pi \Delta x \quad \text{for } n=1, \; m=l, \\
- 2 \pi \Delta x \quad \text{for } 1<n<k, \; m=l, \\
- \frac{3 \pi \Delta x}{2} \quad \text{for } n=k, \; m=l, \\
- \frac{\pi \Delta x}{2} \quad \text{for } n=k+1, \; m=l, \\
0 \quad \text{otherwise}.
\end{cases}, \\
\pardiff{\mathbf{F}_{8(l)}}{\phi_{(1,m)}} &= 0, \\
\pardiff{\mathbf{F}_{8(l)}}{\phi_{x(n,m)}} &= 0, \\
\pardiff{\mathbf{F}_{8(l)}}{\zeta_{(1,m)}} &= 0, \\
\pardiff{\mathbf{F}_{8(l)}}{\zeta_{x(n,m)}} &= 0, \\
\pardiff{\mathbf{F}_{8(l)}}{\psi_{(1,m)}} &= n \quad \text{for } m=l, \\
\pardiff{\mathbf{F}_{8(l)}}{\psi_{x(1,m)}} &= \begin{cases}
x_{1} \quad \text{for } m=l, \\
0 \quad \text{otherwise}.
\end{cases}, \\
\pardiff{\mathbf{F}_{9(l)}}{\psi_{(1,l)}} &= 0, \\
\pardiff{\mathbf{F}_{9(l)}}{\psi_{x(n,m)}} &= \begin{cases}
n - \frac{x_{1}}{\Delta x} \quad \text{for } n=1, \; m=l, \\
\frac{x_{1}}{\Delta x} \quad \text{for } n=2, \; m=l, \\
0 \quad \text{otherwise}.
\end{cases}
\end{align}

\bibliographystyle{jfm}

\begin{thebibliography}{50}
	\expandafter\ifx\csname natexlab\endcsname\relax\def\natexlab#1{#1}\fi
	
	\bibitem[Binder {\em et~al.\/}(2014)Binder, Blyth \& Balasuriya]{Binder2014}
	{\sc Binder, B.~J., Blyth, M.~G. \& Balasuriya, S.} 2014 {Non-uniqueness of
		steady free-surface flow at critical Froude number}. {\em Europhysics
		Letters\/} {\bf 105}, 44003.
	
	\bibitem[Binder {\em et~al.\/}(2013)Binder, Blyth \& McCue]{Binder2013}
	{\sc Binder, B.~J., Blyth, M.~G. \& McCue, S.~W.} 2013 {Free-surface flow past
		arbitrary topography and an inverse approach for wave-free solutions}. {\em
		IMA Journal of Applied Mathematics\/} {\bf 78}, 685--696.
	
	\bibitem[Binder {\em et~al.\/}(2006)Binder, Dias \& Vanden-Broeck]{Binder2006}
	{\sc Binder, B.~J., Dias, F. \& Vanden-Broeck, J.-M.} 2006 {Steady free-surface
		flow past an uneven channel bottom}. {\em Theoretical and Computational Fluid
		Dynamics\/} {\bf 20}, 125--144.
	
	\bibitem[Broutman {\em et~al.\/}(2010)Broutman, Rottman \&
	Eckermann]{Broutman2003}
	{\sc Broutman, D., Rottman, J.~W. \& Eckermann, S.~D.} 2010 A simplified
	{F}ourier method for nonhydrostatic mountain waves. {\em Journal of the
		Atmospheric Sciences\/} {\bf 60}, 2686--2696.
	
	\bibitem[Brown \& Saad(1990)]{Brown1990}
	{\sc Brown, P.~N. \& Saad, Y.} 1990 {Hybrid Krylov methods for nonlinear
		systems of equations}. {\em SIAM Journal on Scientific and Statistical
		Computing\/} {\bf 11}, 450--481.
	
	\bibitem[Chapman \& Vanden-Broeck(2006)]{Chapman2006}
	{\sc Chapman, S.~J. \& Vanden-Broeck, J.-M.} 2006 {Exponential asymptotics and
		gravity waves}. {\em Journal of Fluid Mechanics\/} {\bf 567}, 299--326.
	
	\bibitem[Chardard {\em et~al.\/}(2011)Chardard, Dias, Nguyen \&
	Vanden-Broeck]{Chardard2011}
	{\sc Chardard, F., Dias, F., Nguyen, H.~Y. \& Vanden-Broeck, J.-M.} 2011
	{Stability of some stationary solutions to the forced KdV equation with one
		or two bumps}. {\em Journal of Engineering Mathematics\/} {\bf 70}, 175--189.
	
	\bibitem[Chuang(2000)]{Chuang2000}
	{\sc Chuang, J.~M.} 2000 {Numerical studies on non-linear free surface flow
		using generalized Schwarz-Christoffel transformation}. {\em International
		Journal for Numerical Methods in Fluids\/} {\bf 32}, 745--772.
	
	\bibitem[Darmon {\em et~al.\/}(2014)Darmon, Benzaquen \& Rapha\"{e}l]{darmon14}
	{\sc Darmon, A., Benzaquen, M. \& Rapha\"{e}l, E.} 2014 {Kelvin wake pattern at
		large Froude numbers}. {\em Journal of Fluid Mechanics\/} {\bf 738}, R3.
	
	\bibitem[Dias(2014)]{dias14}
	{\sc Dias, F.} 2014 {Ship waves and Kelvin}. {\em Journal of Fluid Mechanics\/}
	{\bf 746}, 1--4.
	
	\bibitem[Dias \& Vanden-Broeck(1989)]{Dias1989}
	{\sc Dias, F. \& Vanden-Broeck, J.-M.} 1989 {Open channel flows with submerged
		obstructions}. {\em Journal of Fluid Mechanics\/} {\bf 206}, 155--170.
	
	\bibitem[Dias \& Vanden-Broeck(2002)]{Dias2002}
	{\sc Dias, F. \& Vanden-Broeck, J.-M.} 2002 {Generalised critical free-surface
		flows}. {\em Journal of Engineering Mathematics\/} {\bf 42}, 291--301.
	
	\bibitem[Eckermann {\em et~al.\/}(2010)Eckermann, Lindeman, Broutman, Ma \&
	Boybeyi]{Eckermann2010}
	{\sc Eckermann, S.~D., Lindeman, J., Broutman, D., Ma, J. \& Boybeyi, Z.} 2010
	Momentum fluxes of gravity waves generated by variable {F}roude number flow
	over three-dimensional obstacles. {\em Journal of the Atmospheric Sciences\/}
	{\bf 67}, 2260--2278.
	
	\bibitem[Ellingsen(2014)]{ellingsen14}
	{\sc Ellingsen, S.~\r{A}.} 2014 Ship waves in the presence of uniform
	vorticity. {\em Journal of Fluid Mechanics\/} {\bf 742}, R2.
	
	\bibitem[Forbes(1985)]{Forbes1985}
	{\sc Forbes, L.~K.} 1985 On the effects of non-linearity in free-surface flow about a submerged point vortex.  {\em Journal of Engineering Mathematics\/} {\bf 19}, 139--155.
	
	\bibitem[Forbes(1989)]{Forbes1}
	{\sc Forbes, L.~K.} 1989 An algorithm for 3-dimensional free-surface problems
	in hydrodynamics. {\em Journal of Computational Physics\/} {\bf 82},
	330--347.
	
	\bibitem[Forbes \& Hocking(2005)]{Forbes2005}
	{\sc Forbes, L.~K. \& Hocking, G.~C.} 2005 {Flow due to a sink near a vertical
		wall, in infinitely deep fluid}. {\em Computers \& Fluids\/} {\bf 34},
	684--704.
	
	\bibitem[Forbes \& Schwartz(1982)]{Forbes1982}
	{\sc Forbes, L.~K. \& Schwartz, L.~W.} 1982 {Free-surface flow over a
		semicircular obstruction}. {\em Journal of Fluid Mechanics\/} {\bf 114},
	299--314.
	
	\bibitem[Gazdar(1973)]{Gazdar1973}
	{\sc Gazdar, A.~S.} 1973 Generation of waves of small amplitude by an obstacle
	placed on the bottom of a running stream. {\em Journal of the Physical
		Society of Japan\/} {\bf 34}, 530--538.
	
	\bibitem[He {\em et~al.\/}(2014)He, Zhang, Zhu, Wu, Yang, Noblesse, Gu \&
	Li]{he14}
	{\sc He, J., Zhang, C., Zhu, Y., Wu, H., Yang, C.-J., Noblesse, F., Gu, X. \&
		Li, W.} 2014 {Comparison of three simple models of Kelvin's ship wake}. {\em
		European Journal Mechanics - B/Fluids\/} {\bf 49}, 12--19.
	
	\bibitem[Higgins {\em et~al.\/}(2006)Higgins, Read \& Belward]{Higgins2006}
	{\sc Higgins, P.~J., Read, W.~W. \& Belward, S.~R.} 2006 {A series-solution
		method for free-boundary problems arising from flow over topography}. {\em
		Journal of Engineering Mathematics\/} {\bf 54}, 345--358.
	
	\bibitem[Hindmarsh {\em et~al.\/}(2005)Hindmarsh, Brown, Grant, Lee, Serban,
	Shumaker \& Woodward]{Hindmarsh2005}
	{\sc Hindmarsh, A.~C., Brown, P.~N., Grant, K.~E., Lee, S.~L., Serban, R.,
		Shumaker, D.~E. \& Woodward, C.~S.} 2005 {SUNDIALS: Suite of Nonlinear and
		Differential/Algebraic Equation Solvers}. {\em ACM Transactions on
		Mathematical Software\/} {\bf 31}, 363--396.
	
	\bibitem[Hocking {\em et~al.\/}(2013)Hocking, Holmes \& Forbes]{Hocking2013}
	{\sc Hocking, G.~C., Holmes, R.~J. \& Forbes, L.~K.} 2013 {A note on waveless
		subcritical flow past a submerged semi-ellipse}. {\em Journal of Engineering
		Mathematics\/} {\bf 81}, 1--8.
	
	\bibitem[King \& Bloor(1990)]{King1990}
	{\sc King, A.~C. \& Bloor, M. I.~G.} 1990 {Free-surface flow of a stream
		obstructed by an arbitrary bed topography}. {\em Quarterly Journal of
		Mechanics and Applied Mathematics\/} {\bf 43}, 87--106.
	
	\bibitem[Knoll \& Keyes(2004)]{Knoll2004}
	{\sc Knoll, D.~A. \& Keyes, D.~E.} 2004 {Jacobian-free Newton--Krylov methods:
		a survey of approaches and applications}. {\em Journal of Computational
		Physics\/} {\bf 193}, 357--397.
	
	\bibitem[Lamb(1932)]{Lamb1932}
	{\sc Lamb, H.} 1932 {\em {Hydrodynamics}\/}. Cambridge University Press.
	
	\bibitem[Li \& Ellingsen(2016)]{li16}
	{\sc Li, Y. \& Ellingsen, S.~\r{A}.} 2016 Ship waves on uniform shear current
	at finite depth: wave resistance and critical velocity. {\em Journal Fluid
		Mechanics\/} {\bf 791}, 539--567.
	
	\bibitem[Lustri {\em et~al.\/}(2012)Lustri, McCue \& Binder]{Lustri2012}
	{\sc Lustri, C.~J., McCue, S.~W. \& Binder, B.~J.} 2012 {Free surface flow past
		topography: a beyond-all-orders approach}. {\em European Journal of Applied
		Mathematics\/} {\bf 23}, 441--467.
	
	\bibitem[McCue \& Forbes(2002)]{mccue2002}
	{\sc McCue, S.~W. \& Forbes, L.~K.} 2002 Free-surface flows emerging from beneath a semi-infinite plate with constant vorticity.  {\em Journal of Fluid Mechanics\/} {\bf 461}, 387--407.
	
	\bibitem[Miao \& Liu(2015)]{miao15}
	{\sc Miao, S. \& Liu, Y.} 2015 Wave pattern in the wake of an arbitrary moving
	surface pressure disturbance. {\em Physics of Fluids\/} {\bf 27}, 122102.
	
	\bibitem[Noblesse {\em et~al.\/}(2014)Noblesse, He, Zhu, Hong, Zhang, Zhu \&
	Yang]{noblesse14}
	{\sc Noblesse, F., He, J., Zhu, Y., Hong, L., Zhang, C., Zhu, R. \& Yang, C.}
	2014 {Why can ship wakes appear narrower than Kelvin's angle?} {\em European
		Journal of Mechanics - B/Fluids\/} {\bf 46}, 164--171.
	
	\bibitem[P{\u{a}}r{\u{a}}u \& Vanden-Broeck(2002)]{Parau1}
	{\sc P{\u{a}}r{\u{a}}u, E.~I. \& Vanden-Broeck, J.-M.} 2002 Nonlinear two-and
	three-dimensional free surface flows due to moving disturbances. {\em
		European Journal of Mechanics-B/Fluids\/} {\bf 21}, 643--656.
	
	\bibitem[P{\u{a}}r{\u{a}}u \& Vanden-Broeck(2011)]{Parau2011}
	{\sc P{\u{a}}r{\u{a}}u, E.~I. \& Vanden-Broeck, J.-M.} 2011 {Three-dimensional
		waves beneath an ice sheet due to a steadily moving pressure}. {\em
		Philosophical Transactions of the Royal Society of London A: Mathematical,
		Physical and Engineering Sciences\/} {\bf 369}, 2973--2988.
	
	\bibitem[P{\u{a}}r{\u{a}}u {\em et~al.\/}(2005{\natexlab{{\em
				a\/}}})P{\u{a}}r{\u{a}}u, Vanden-Broeck \& Cooker]{Parau2005}
	{\sc P{\u{a}}r{\u{a}}u, E.~I., Vanden-Broeck, J.-M. \& Cooker, M.~J.}
	2005{\natexlab{{\em a\/}}} {Nonlinear three-dimensional gravity--capillary
		solitary waves}. {\em Journal of Fluid Mechanics\/} {\bf 536}, 99--105.
	
	\bibitem[P{\u{a}}r{\u{a}}u {\em et~al.\/}(2005{\natexlab{{\em
				b\/}}})P{\u{a}}r{\u{a}}u, Vanden-Broeck \& Cooker]{Parau2}
	{\sc P{\u{a}}r{\u{a}}u, E.~I., Vanden-Broeck, J.-M. \& Cooker, M.~J.}
	2005{\natexlab{{\em b\/}}} Three-dimensional gravity-capillary solitary waves
	in water of finite depth and related problems. {\em Physics of Fluids\/} {\bf
		17}, 122101.
	
	\bibitem[P{\u{a}}r{\u{a}}u {\em et~al.\/}(2007{\natexlab{{\em a\/}}})P{\u{a}}r{\u{a}}u, Vanden-Broeck \& Cooker]{Parau2007a}
	{\sc P{\u{a}}r{\u{a}}u, E.~I., Vanden-Broeck, J.-M. \& Cooker, M.~J.} 2007{\natexlab{{\em
				a\/}}}  {Nonlinear three-dimensional interfacial flows with a free surface}. {\em
		Journal of Fluid Mechanics\/} {\bf 591}, 481--494.
	
	\bibitem[P{\u{a}}r{\u{a}}u {\em et~al.\/}(2007{\natexlab{{\em b\/}}})P{\u{a}}r{\u{a}}u, Vanden-Broeck \& Cooker]{Parau2007b}
	{\sc P{\u{a}}r{\u{a}}u, E.~I., Vanden-Broeck, J.-M. \& Cooker, M.~J.} 2007{\natexlab{{\em
				b\/}}}  {Three-dimensional capillary-gravity waves generated by a moving disturbance}. {\em
		Physics of Fluids\/} {\bf 19}, 082102.
	
	\bibitem[Pethiyagoda {\em et~al.\/}(2014{\natexlab{{\em a\/}}})Pethiyagoda,
	McCue, Moroney \& Back]{Pethiyagoda14}
	{\sc Pethiyagoda, R., McCue, S.~W., Moroney, T.~J. \& Back, J.~M.}
	2014{\natexlab{{\em a\/}}} {Jacobian-free Newton--Krylov methods with GPU
		acceleration for computing nonlinear ship wave patterns}. {\em Journal of
		Computational Physics\/} {\bf 269}, 297--313.
	
	\bibitem[Pethiyagoda {\em et~al.\/}(2014{\natexlab{{\em b\/}}})Pethiyagoda,
	McCue \& Moroney]{pethiyagoda2014}
	{\sc Pethiyagoda, R., McCue, S.~W. \& Moroney, T.~J.} 2014{\natexlab{{\em
				b\/}}} What is the apparent angle of a Kelvin ship wave pattern? {\em Journal
		of Fluid Mechanics\/} {\bf 758}, 468--485.
	
	\bibitem[Pethiyagoda {\em et~al.\/}(2015)Pethiyagoda, McCue \&
	Moroney]{pethiyagoda15}
	{\sc Pethiyagoda, R., McCue, S.~W. \& Moroney, T.~J.} 2015 Wake angle for
	surface gravity waves on a finite depth fluid. {\em Physics of Fluids\/} {\bf
		27}, 061701.
	
	\bibitem[Pethiyagoda {\em et~al.\/}(2017)Pethiyagoda, McCue \&
	Moroney]{pethiyagoda17}
	{\sc Pethiyagoda, R., McCue, S.~W. \& Moroney, T.~J.} 2017 Spectrograms of ship
	wakes: identifying linear and nonlinear wave signals. {\em Journal of Fluid
		Mechanics\/} {\bf 811}, 189--209.
	
	
	\bibitem[Pethiyagoda {\em et~al.\/}(2018{\natexlab{{\em a\/}}})Pethiyagoda,
	Moroney, Macfarlane, Binns \& McCue]{pethiyagoda17b}
	{\sc Pethiyagoda, R., Moroney, T.~J., Macfarlane, G.~J., Binns, J.~R. \& McCue, S.~W.} 2018{\natexlab{{\em a\/}}} Time-frequency analysis of ship wave patterns in shallow water: modelling and experiments.  {\em Ocean Engineering}, in press,   doi:10.1016/j.oceaneng.2018.01.108.
	
	
	\bibitem[Pethiyagoda {\em et~al.\/}(2018{\natexlab{{\em b\/}}})Pethiyagoda, Moroney  \&
	McCue]{pethiyagoda17c}
	{\sc Pethiyagoda, R., Moroney, T.~J. \& McCue, S.~W.} 2018{\natexlab{{\em b\/}}} Efficient computation of two-dimensional steady free-surface flows. {\em International Journal for Numerical Methods in Fluids\/} {\bf 86}, 607--624.
	
	
	\bibitem[Rabaud \& Moisy(2013)]{rabaud13}
	{\sc Rabaud, M. \& Moisy, F.} 2013 {Ship wakes: Kelvin or Mach angle?} {\em
		Physical Review Letters\/} {\bf 110}, 214503.
	
	\bibitem[Saad \& Schultz(1986)]{Saad1986}
	{\sc Saad, Y. \& Schultz, M.~H.} 1986 {GMRES: A generalized minimal residual
		algorithm for solving nonsymmetric linear systems}. {\em SIAM Journal on
		Scientific and Statistical Computing\/} {\bf 7}, 856--869.
	
	\bibitem[Scullen(1998)]{Scullen98}
	{\sc Scullen, D.~C.} 1998 {Accurate computation of steady nonlinear
		free-surface flows}. PhD Thesis, Department of Applied Mathematics, University of Adelaide.
	
	\bibitem[Smeltzer \& Ellingsen(2017)]{smeltzer17}
	{\sc Smeltzer, B.~K. \& Ellingsen, S.~\r{A}.} 2017 Surface waves on currents
	with arbitrary vertical shear. {\em Physics of Fluids\/} {\bf 29}, 047102.
	
	\bibitem[Soomere(2007)]{soomere2007}
	{\sc Soomere, T.} 2007 Nonlinear components of ship wake waves.  {\em Applied Mechanics Reviews\/} {\bf 60}, 120--138.
	
	\bibitem[Teixeira(2014)]{Teixeira2014}
	{\sc Teixeira, M. A.~C.} 2014 The physics of orographic gravity wave drag. {\em
		Frontiers in Physics\/} {\bf 2}, 43.
	
	\bibitem[Torsvik {\em et~al.\/}(2015)Torsvik, Soomere, Didenkulova \&
	Sheremet]{torsvik15a}
	{\sc Torsvik, T., Soomere, T., Didenkulova, I. \& Sheremet, A.} 2015
	Identification of ship wake structures by a time-frequency method. {\em
		Journal of Fluid Mechanics\/} {\bf 765}, 229--251.
	
	\bibitem[Wade {\em et~al.\/}(2017)Wade, Binder, Mattner \& Denier]{Wade2017}
	{\sc Wade, S.~L., Binder, B.~J., Mattner, T.~W. \& Denier, J.~P.} 2017 {Steep
		waves in free-surface flow past narrow topography}. {\em Physics of Fluids\/}
	{\bf 29}, 062107.
	
	\bibitem[Wehausen \& Laitone(1960)]{wehausen60}
	{\sc Wehausen, J.~V. \& Laitone, E.~V.} 1960 {\em Surface waves\/}. Springer.
	
	\bibitem[Zhang {\em et~al.\/}(2015)Zhang, He, Zhu, Yang, Li, Zhu, Lin \&
	Noblesse]{zhang15a}
	{\sc Zhang, C., He, J., Zhu, Y., Yang, C.-J., Li, W., Zhu, Y., Lin, M. \&
		Noblesse, F.} 2015 Interference effects on the Kelvin wake of a monohull ship
	represented via a continuous distribution of sources. {\em European Journal
		of Mechanics - B/Fluids\/} {\bf 51}, 27--36.
	
	\bibitem[Zhang \& Zhu(1996{\natexlab{{\em a\/}}})]{Zhang1996b}
	{\sc Zhang, Y. \& Zhu, S.} 1996{\natexlab{{\em a\/}}} A comparison study of
	nonlinear waves generated behind a semicircular trench. {\em Proceedings of
		the Royal Society of London A\/} {\bf 452}, 1563--1584.
	
	\bibitem[Zhang \& Zhu(1996{\natexlab{{\em b\/}}})]{Zhang1996}
	{\sc Zhang, Y. \& Zhu, S.} 1996{\natexlab{{\em b\/}}} {Open channel flow past a
		bottom obstruction}. {\em Journal of Engineering Mathematics\/} {\bf 30},
	487--499.
	
	\bibitem[Zhu {\em et~al.\/}(2015)Zhu, He, Zhang, Wu, Wan, Zhu \&
	Noblesse]{zhu15}
	{\sc Zhu, Y., He, J., Zhang, C., Wu, H., Wan, D., Zhu, R. \& Noblesse, F.} 2015
	Farfield waves created by a monohull ship in shallow water. {\em European
		Journal of Mechanics - B/Fluids\/} {\bf 49}, 226--234.
	
\end{thebibliography}

\end{document}